%% file: main.tex
\documentclass[10pt]{iopart}
\input{packages_commands}
\usepackage{esint}
\usepackage{amsmath}

\hypersetup{
     colorlinks   = true,
     linkcolor    = blue,
     citecolor    = blue,
     urlcolor     = blue
}

\begin{document}
\title{First 2D electron density measurements using Coherence Imaging Spectroscopy in the MAST-U Super-X divertor
}
\author{N. Lonigro$^1$, R. Doyle$^{2,3}$, J. S. Allcock$^2$, B. Lipschultz$^1$, K.Verhaegh$^2$, C. Bowman$^2$, D. Brida$^4$, J. Harrison$^2$, O. Myatra$^2$, S. Silburn$^2$,C. Theiler$^5$,  T.A. Wijkamp$^{6,7}$, MAST-U Team$^8$ and the EUROfusion Tokamak Exploitation Team$^9$}
\address{$^1$York Plasma Institute, University of York, United Kingdom  }
\address{$^2$UKAEA, Culham Campus, United Kingdom }
\address{$^3$NCPST, Dublin City University, Ireland}
\address{$^4$ Max Planck Institute for Plasma Physics, Garching, Germany}  
\address{$^5$5 Ecole Polytechnique Federale de Lausanne (EPFL), Swiss Plasma Center (SPC), Switzerland}
\address{$^6$Department of Applied Physics, Eindhoven University of Technology, Netherlands} 
\address{$^7$DIFFER, Eindhoven, Netherlands}
\address{$^8$See the author list of J. Harrison et al 2019 Nucl. Fusion 59 112011}
\address{$^9$See the author list of “Progress on an exhaust solution for a reactor using EUROfusion multi-machines capabilities” by E. Joffrin et al. to
be published in Nuclear Fusion Special Issue: Overview and Summary Papers from the 29th Fusion Energy Conference 2023}

\ead{\mailto{nicola.lonigro@ukaea.uk}}

\ioptwocol
\begin{abstract}
 2D profiles of electron density and neutral temperature are inferred from multi-delay Coherence Imaging Spectroscopy data of divertor plasmas using a non-linear inversion technique. The inference is based on imaging the spectral line-broadening of Balmer lines and can differentiate between the Doppler and Stark broadening components by measuring the fringe contrast at multiple interferometric delays simultaneously. The model has been applied to images generated from simulated density profiles to evaluate its performance. Typical mean absolute errors of 30 \% are achieved, which are consistent with Monte Carlo uncertainty propagation accounting for noise, uncertainties in the calibrations, and in the model inputs. The analysis has been tested on experimental data from the MAST-U Super-X divertor, where it infers typical electron densities of 2-3 $10^{19}$ m$^{-3}$ and neutral temperatures of 0-2 eV during beam-heated L-mode discharges. The results are shown to be in reasonable agreement with the other available diagnostics.     
\end{abstract}


\section{Introduction}
Mitigating divertor target damage from the impinging power and particle fluxes is one of the main challenges on the path to commercial fusion reactors. Alternative divertor concepts\cite{soukhanovskii_review_2017} are being studied on a variety of devices, such as TCV \cite{fil_TCV_2020}\cite{maurizio_TCV_2018}, DIID\cite{soukhanovskii_DIIID_2015} and MAST-U \cite{verhaegh_role_2023}\cite{verhaegh_spectroscopic_2022}, as a mitigation strategy in case more conventional designs prove to be unsuitable for future devices. Advanced diagnostics are required to characterize and optimize the performance of these different divertor concepts, and imaging diagnostics have proven themselves as a powerful tool for quantitative 2D measurements in the divertor \cite{perek_emissivity_TCV}\cite{wijkamp_MWI_2023}\cite{linehan_validation_2023}. Inferring plasma parameters such as electron density and temperature by modeling Balmer line emissivities, or their ratios, can be complicated by the presence of molecular effects, which may significantly affect the measured emissivities \cite{verhaegh_molecules_1_2021}\cite{verhaegh_molecules_2_2021}\cite{verhaegh_role_2023}. Spectrometers have been long used to infer emission-weighted, line-averaged electron densities in divertors by analysing the Stark broadening of the hydrogen emission spectra \cite{lomanowski_Stark_fit_2015}. However, this has the disadvantage of only providing (emission-weighted) line-averaged measurements along a limited number of lines of sight. Coherence Imaging Spectroscopy (CIS) provides a way to combine the advantages of these two measurement techniques by providing direct 2D density measurements based on Stark broadening, thus unaffected by emissivity inference uncertainties.    
It has been used to determine the electron density in the assumption of a purely Stark broadened line on the linear device Pilot-PSI \cite{lischtschenko_density_2010}. Using a multi-delay configuration, the density was then estimated while accounting for both Stark and Doppler broadening in Magnum-PSI \cite{Allcock_CIS_density}. This work aims to extend the previous results to a tokamak for the first time and present an inversion technique to infer 2D poloidal profiles of electron density and neutral temperature in the MAST-U Super-X divertor. The instrument used in this work is a multi-delay CIS system installed as the last channel of the Multi-Wavelength Imaging diagnostic (MWI), a set of 11 cameras sharing the same view of the MAST-U lower divertor based on a polychromator design \cite{feng_development_2021}\cite{Doyle_CIS}. 
\section{Coherence Imaging Spectroscopy}
Coherence Imaging Spectroscopy \cite{Howard_Overview_CIS} is a diagnostic technique based on Fourier Transform Spectroscopy. It encodes information about the emission spectrum of narrow-band spectra in an interference pattern overlaid on the image of the plasma emission. The plasma parameters can then be inferred by modeling their relation to the spectrum and consequently to the observed pattern.\\
In current applications, the interference pattern is generated by a polarization interferometer that splits the incoming light along two polarization states and applies a phase delay between them, before they interfere at the camera sensor.
For a given area-normalized spectrum $g(\nu)$ of a spectral line centered at frequency $\nu_0$, the light reaching the sensor after being delayed by an interferometric delay $\phi_0$ will be characterized by a complex coherence value $\gamma(\phi_0)$ \cite{Allcock_CIS_density}

\begin{equation}
    \gamma(\phi_0)\approx \int^\infty_{-\infty}g(\nu)\exp \left\{ i \phi_0 \left[1 + \kappa_0 \left(\frac{\nu-\nu_0}{\nu_0} \right) \right]\right\} \dd \nu
    \label{eq:gamma}
\end{equation}

where $\kappa_0$ is a first-order approximation of the light dispersion through the crystals and can be measured during the characterization of the instrument. The modulus and argument of $\gamma$ will be the contrast and phase of the fringe pattern respectively

\begin{align}
    \zeta = & \abs{\gamma} & \Phi = \arg \gamma
    \label{eq:contrast}
\end{align}

An example of the modeled interference pattern corresponding to a homogeneous plasma source is shown in figure \ref{fig:fig1_cell/interferogram}.  

The measured interference pattern can be demodulated into three components:

\begin{itemize}
    \item a DC image, which is equivalent to the brightness image from a conventional, non-polarized, camera.
    \item  a contrast image, which is a diagnostic for changes in the spectral line width and shape.
    \item  a phase image, which is a sensitive diagnostic for changes in the spectral line centre wavelength.
\end{itemize}
 The phase information has been used to measure the impurity ion velocity on MAST \cite{Silburn_CIS_MAST}, DIIID \cite{Weber_CIS_DIIID}\cite{Allen_CIS_DIIID} and W7-X \cite{Perseo_W7X_CIS}\cite{Perseo_W7X_CIS_2}, as well as the neutral flow in ASDEX-U \cite{Gradic_CIS_ASDEX}, by modeling the effect of Doppler shifts on the observed lineshape. A CIS diagnostic is also planned for ITER \cite{howard_CIS_ITER}. The contrast of the fringe pattern is instead a measure of the broadening of the line, as a more broadened spectrum will lead to less coherent light reaching the camera and so a less visible interference pattern. It has been used in ASDEX-U to determine the impurity ion temperature through its Doppler broadening of the CIII line\cite{Gradic_CIS_W7X}. 
A method for the simultaneous inversion of impurity velocity and ion temperature is under development at MAST-U \cite{Doyle_inversion}. 

\subsection{Multi-delay Coherence Imaging Spectroscopy}

\begin{figure}[h!]\centering
\vspace{-0.0 cm} 
\includegraphics[width= 0.5\textwidth]{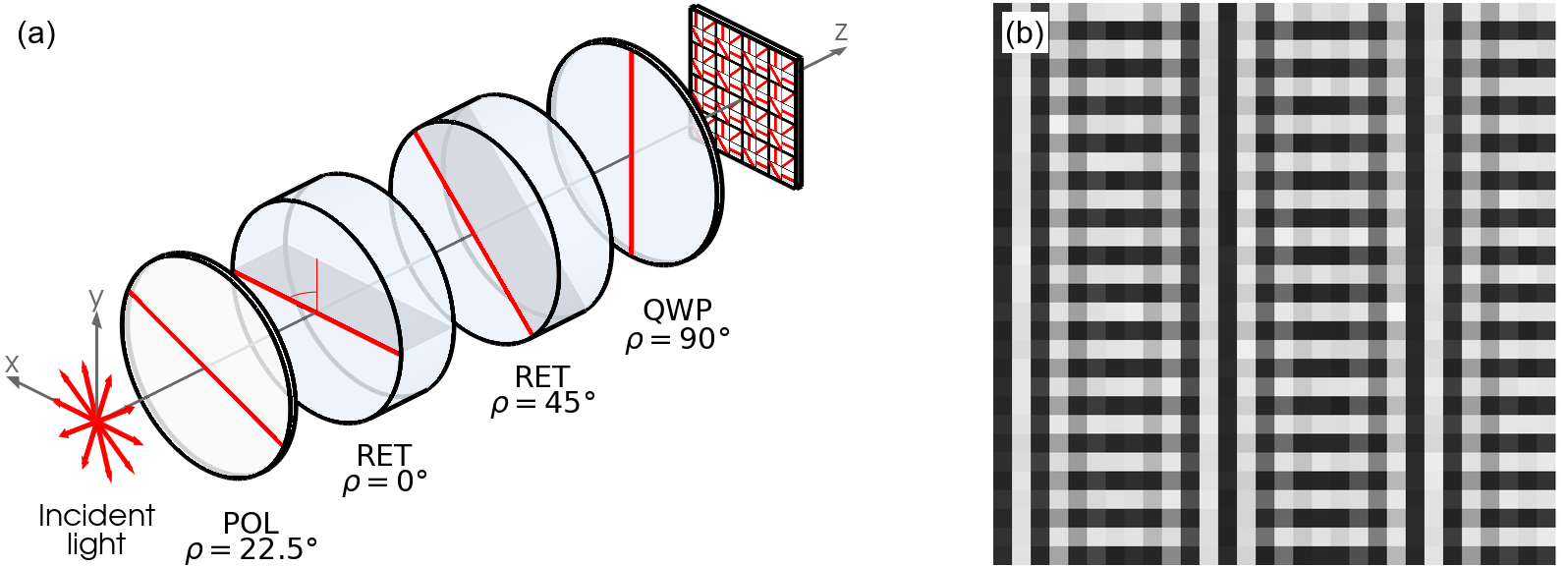}
\vspace{-0.5 cm}\caption{(a) MAST-U CIS interferometer cell made up of a polariser, two retarding crystals (displacer plate and waveplate), a quarter wave plate and a polarized camera sensor. (b) Example CIS interferogram for a homogeneous plasma source}
\label{fig:fig1_cell/interferogram}
\vspace{-0.3cm} 
\end{figure}

To diagnose more complex lineshapes, the MAST-U CIS diagnostic is designed to overlay three interference patterns on the same image, each corresponding to a different interferometric delay. This allows disentangling multiple broadening mechanisms by exploiting their different dependence on the delay $\phi_0$ in equation(\ref{eq:gamma})\cite{Allcock_CIS_density}. The layout of the MAST-U polarization interferometer is shown in figure \ref{fig:fig1_cell/interferogram} and it uses two birefringent crystals as the two retarders.  

The setup is composed of a polarizer, a displacer plate as a first retarder crystal, a waveplate as the second retarder, and a quarter waveplate in front of a polarization sensitive sensor. 

The polarized light from the first crystal is split into two perpendicularly polarized components by the displacer plate, which will be delayed with respect to each other by a phase delay $\phi_D$. After exiting the displacer plate one component will be laterally displaced from the other in a direction that is determined by the crystal optic axis, thus resulting in a sheared delay across the sensor and making it a pixel-dependent quantity $\phi_1(x,y) = \phi_D + \delta\phi(x,y)$. The light will then reach the waveplate, where each component will be further divided into two components, one of which will be delayed by an additional phase $\phi_2$. The quarter waveplate will then collapse all the components together into a sum of circularly polarized waves that will interfere at the sensor.

The expression for the measured signal can be found using Muller calculus as the product of the Muller matrices representing each component and is given by eq (\ref{eq:interferogram})\cite{Allcock_CIS_density} \cite{Doyle_CIS}
\begin{multline}
    I_{out}(x,y) = \frac{I_{in}}{4} \left[ 1 + \frac{\sqrt{2}}{2}\zeta_2\cos\left(\phi_2 + m\frac{\pi}{2}\right) + \right.  \\  + \left. \frac{\sqrt{2}}{4}\zeta_{2-1}\cos\left(\phi_2 +  m\frac{\pi}{2} - \phi_1(x,y) \right) \right. \\  
    \left. - \frac{\sqrt{2}}{4}\zeta_{2+1}\cos\left(\phi_2 + m\frac{\pi}{2} + \phi_1(x,y)\right) \right]
    \label{eq:interferogram}
\end{multline}

where $m \in [0,1,2,3] $ is the index representing the orientation of the polarizer in front of each pixel. The measured interferogram will be the sum of a DC component and three interference patterns corresponding to the interferometric delays of $\phi_2,\phi_2 - \phi_1$ and $\phi_2 + \phi_2$. The contrast measurement for each delay can be obtained using standard 2D Fourier demodulation techniques. Compared to the previous four-delay design \cite{Allcock_CIS_density}, this configuration increases the throughput by removing one of the polarizers at the cost of measuring at 3 delays instead of 4. The delay values of the two crystals used for this work are $\phi_1 \approx 10 \cdot 10^3$ and $\phi_2 \approx 18 \cdot 10^3$ radians at 433.9 nm. More technical details on the diagnostic and its calibration procedures can be found in \cite{Doyle_CIS}.

\section{Lineshape Model}\label{sec:lineshape}
\begin{figure*}[h!]\centering
\begin{subfigure}[b]{0.35\textwidth}
\includegraphics[width= \textwidth]{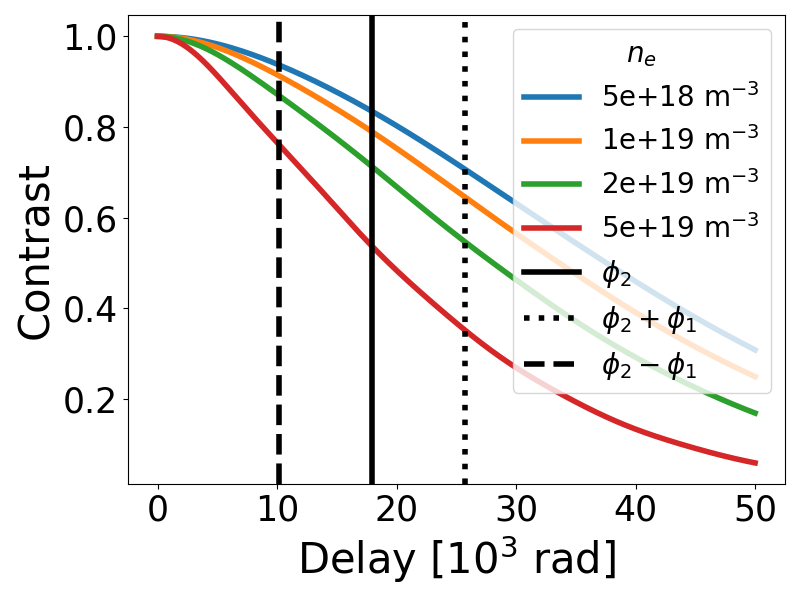}
\caption{Fixed $T_n = 1$ eV}
\end{subfigure}
\begin{subfigure}[b]{0.35\textwidth}
\includegraphics[width= \textwidth]{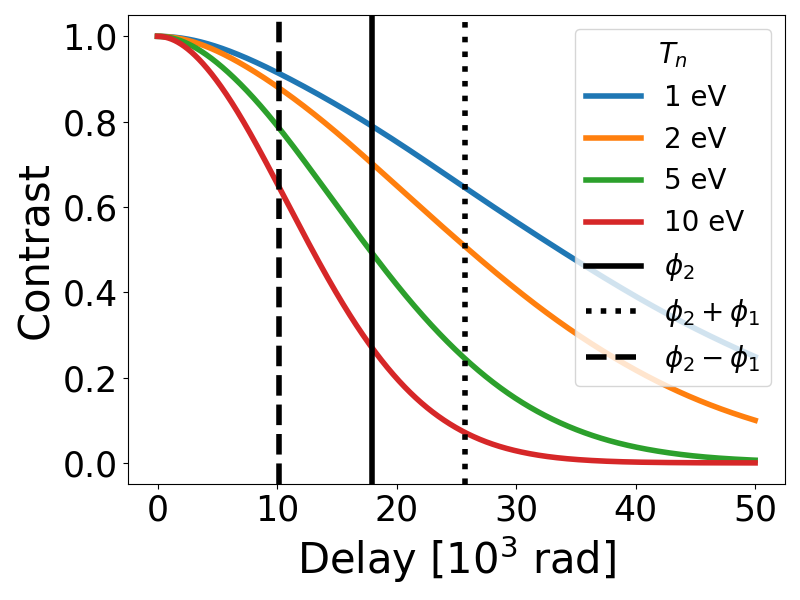}
\caption{Fixed $n_e = 1 \cdot 10^{19}$ m$^{-3}$}
\end{subfigure}
\vspace{-0.3 cm }\caption{Contrast dependence as a function of delay for varying plasma conditions in the absence of background emission. The vertical lines represent the three delay values used in the MAST-U system.}
\label{fig:fig2_delay_curves}
\vspace{-0.3cm} 
\end{figure*}
\begin{figure*}[h!]\centering
\includegraphics[width= 0.49\textwidth]{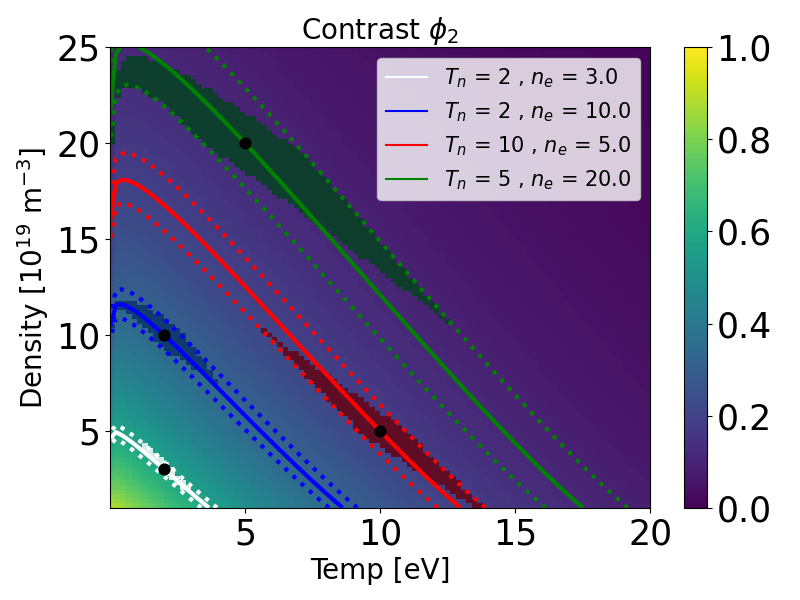}
\includegraphics[width= 0.49\textwidth]{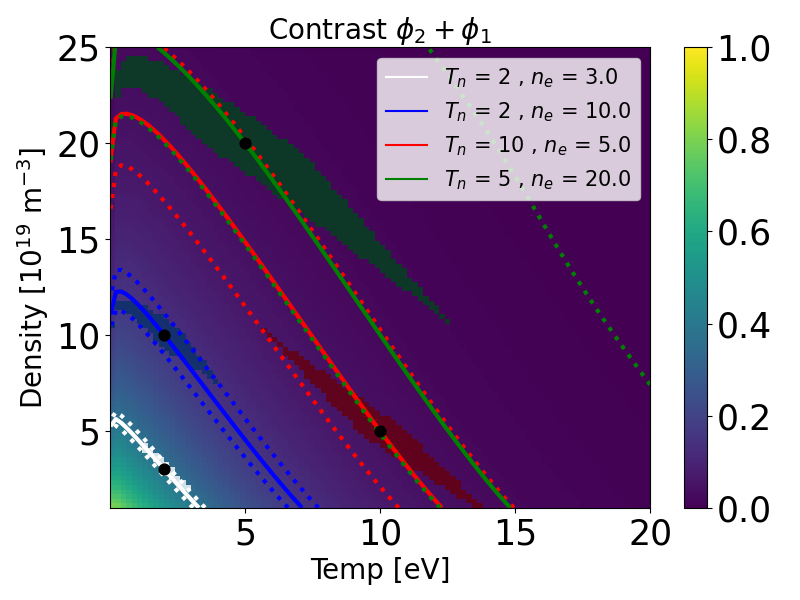}
\includegraphics[width= 0.49\textwidth]{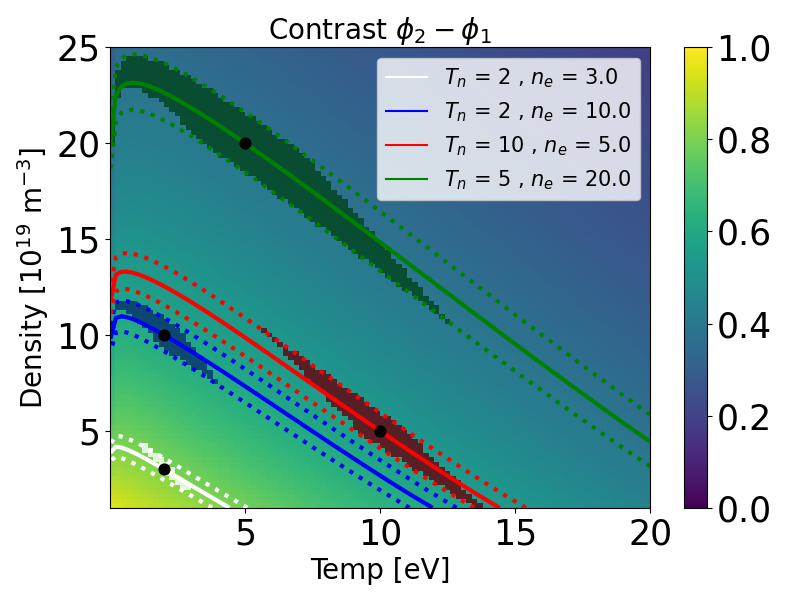}
\vspace{-0.3 cm }\caption{Contrast dependence as a function of neutral temperature and electron density for the three delays used in the MAST-U system. For 4 couples of ($T_n, n_e$) values, the solid line is a contour line for the corresponding contrast value, the region between the dotted lines shows the area of parameter space within uncertainty for that single contrast measurement, while the shaded area represents the region in agreement with all three delay measurements. An uncertainty of 0.02 in contrast on each delay is assumed, in agreement with typical noise values on a single pixel.}
\label{fig:fig2.5_2D_contrast_var}
\vspace{-0.3cm} 
\end{figure*}
Typical electron densities expected in the MAST-U divertor are in the range $n_e \sim 10^{19} - 10^{20}$ m$^{-3}$, while typical neutral temperatures can reach a few eV. In the divertor chamber the magnitude of the magnetic field is lower than 1 T. The $D_\gamma$ spectral line has been chosen as the target for coherence imaging in a trade-off between the higher signal of low $n$ Balmer lines and the stronger Stark broadening of higher $n$ lines. In these conditions, for the $D_\gamma$ spectral line, Doppler and Stark broadening are expected to be dominant  while Zeeman splitting has a minor effect. Doppler broadening can be modeled with a Gaussian lineshape, while the Stark broadening lineshape is approximated using a modified Lorentzian fit \cite{lomanowski_Stark_fit_2015}

\begin{align}
    g_G(\lambda) = & \frac{c}{v_{Th}\lambda_0\sqrt{\pi}} \exp\left(- \left( \frac{ c \left( \lambda_0 -\lambda \right) }{v_{Th} \lambda_0} \right)^2\right)\\
    g_S(\lambda) = & \left [ \left (\lambda - \lambda_0 \right)^{\frac{5}{2}} + \left(\frac{c}{2}n_e^aT_e^{-b}\right)^{\frac{5}{2}} \right]^{-1} \label{eq:stark_profile}
\end{align}
where $v_{Th} = \sqrt{2 T_n k_b/m_D}$ is the thermal velocity of the deuterium neutrals, $\lambda_0 = 433.928$ nm is the central wavelength of the Balmer $D_\gamma$ line and the coefficients of the Stark broadening fit are ($a = 0.6796, b = 0.03, c= 1.31\cdot 10^{-15}$). The dependence of the Stark broadening term on the temperature is very weak, thus the expression can be simplified further by assuming $T_e \approx T_n$. Performing a convolution of the two lineshape profiles, both effects can be taken into account simultaneously.

For a fixed density and temperature values, plugging the expression of the normalized spectrum in equations \ref{eq:gamma} and \ref{eq:contrast} gives the contrast as a function of interferometric delay. The dependence on delay will be significantly different for the two broadening mechanisms due to the different ways in which they affect the lineshape. The Stark term affects mostly the wings of the distribution, thus resulting in a lower slope in delay space. Examples of varying density with fixed temperature and varying temperature with fixed density are shown in figure \ref{fig:fig2_delay_curves}.

The advantage of using multiple delays is highlighted in figure \ref{fig:fig2.5_2D_contrast_var}, where for four examples of ($T_n, n_e$) pairs the region of parameter space compatible with the three measurements is highlighted as a shaded region.   
Compared to the corresponding inference using each delay by itself, the region between the two dotted lines, inferences using all three delay values allow significantly restricting the region of parameter space in agreement with the measurements,  resulting in bounds on each parameter independently of the other. As the neutral temperature increases for fixed density, the uncertainty on the density increases as the Doppler broadening starts to completely dominate the spectrum. Analogously, higher density for the same neutral temperature leads to lower relative uncertainty on the density. As the spectral line becomes broader (through either broadening mechanisms or a combination of both), the relative uncertainty starts rising as the contrast starts decreasing below the sensitive range of the instrument. This is particularly visible in the $\phi_2 + \phi_1$ term for temperature above 10 eV or densities above $2 \cdot 10^{20} m^{-3}$. Tuning for higher densities and temperatures could be achieved by designing an instrument to measure at lower delay values. The uncertainty assumed in the figure \ref{fig:fig2_delay_curves} corresponds to the typical noise in the contrast measured, but it could be reduced further though increased light throughout, downsampling of the image to average multiple pixels or time averaging of multiple frames.
\subsection{Sensitivity studies}

Apart from Stark broadening and Doppler broadening, other effects such as plasma background emission, Zeeman splitting and contaminating impurity transitions can influence the observed lineshape - and thus the CIS measurements. To probe the sensitivity of our measurements to these various additional effects, we will simulate and vary them independently. 

The effect of background emission $C$, such as bremsstrahlung, will be to reduce the coherence of the light and lower the measured contrast. The $D_\gamma$ bandpass filter used in this work has a full-width-half-maximum of $\sim 1.5$nm, much larger than the width of the spectral line. In these conditions and for the three delays used in this work, the background will act as the same multiplicative degrading factor on all three contrast measurements\cite{Allcock_CIS_density}.

\begin{figure*}[h!]\centering
\begin{subfigure}[b]{0.35\textwidth}
\includegraphics[width= \textwidth]{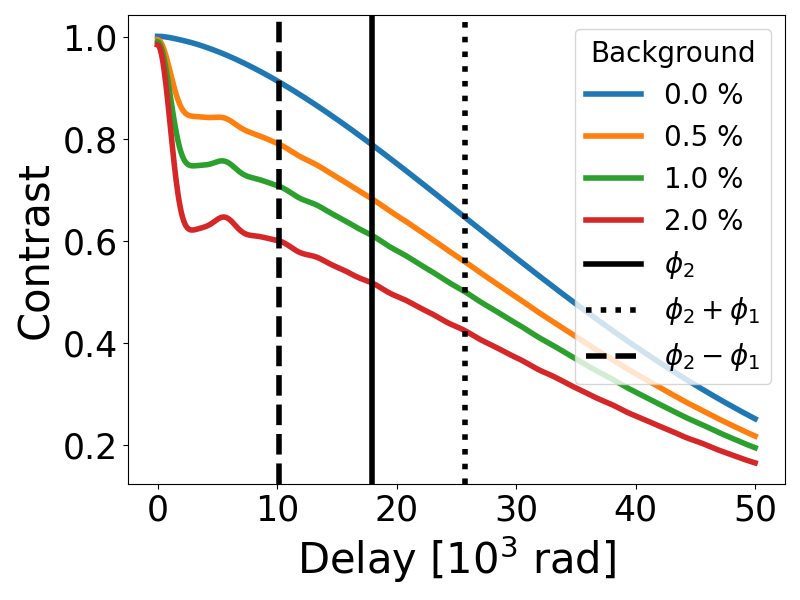}
\caption{}
\end{subfigure}
\begin{subfigure}[b]{0.35\textwidth}
\includegraphics[width=\textwidth]{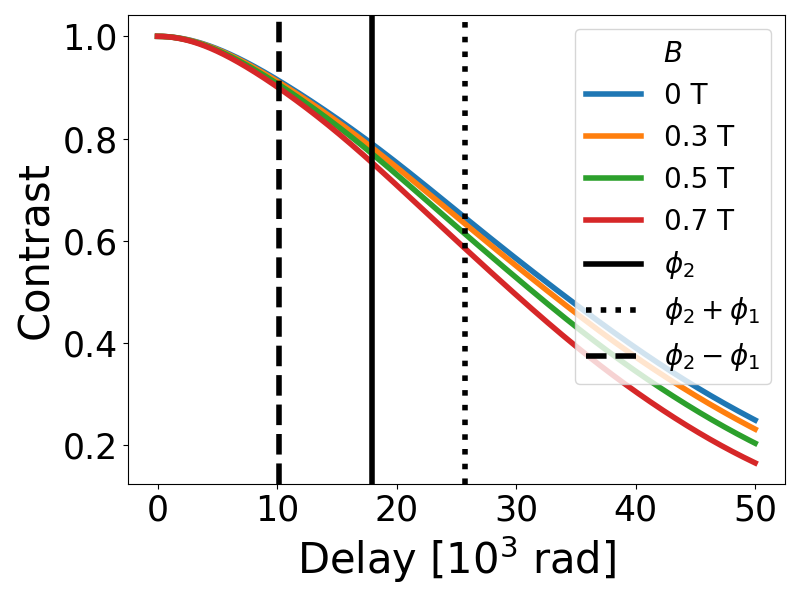}
\caption{}
\end{subfigure}
\vspace{-0.3 cm}\caption{Contrast dependence on phase delay for a plasma with $n_e = 1e19$ m$^{-3}$, $T_n = 1$ eV and (a) varying levels of background emission or (b) varying levels of magnetic field, for a tangential view and a $\pi/4$ angle with the front polarizer.}
\label{fig:fig3_delay_curve_background}
\vspace{-0.3cm} 
\end{figure*}
\begin{figure*}[h!]\centering
\begin{subfigure}[b]{0.35\textwidth}
\includegraphics[width= \textwidth]{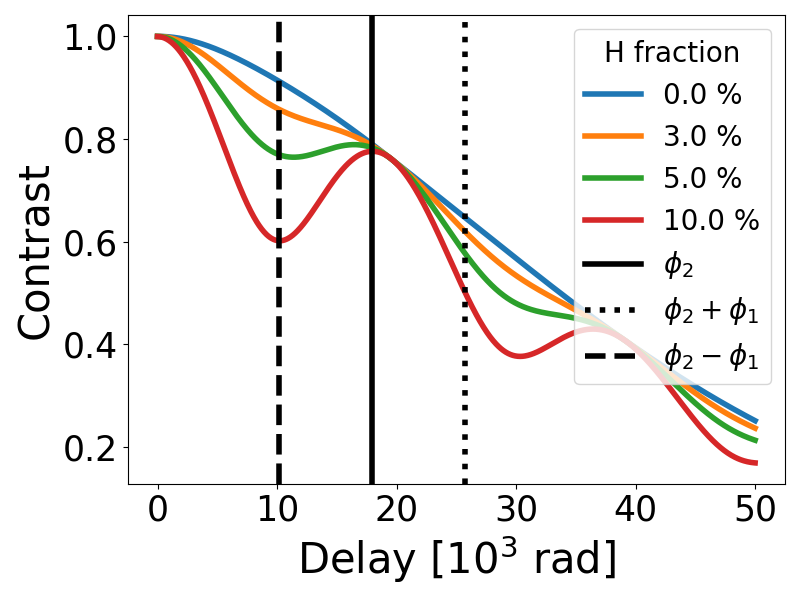}
\caption{}
\end{subfigure}
\begin{subfigure}[b]{0.35\textwidth}
\includegraphics[width=\textwidth]{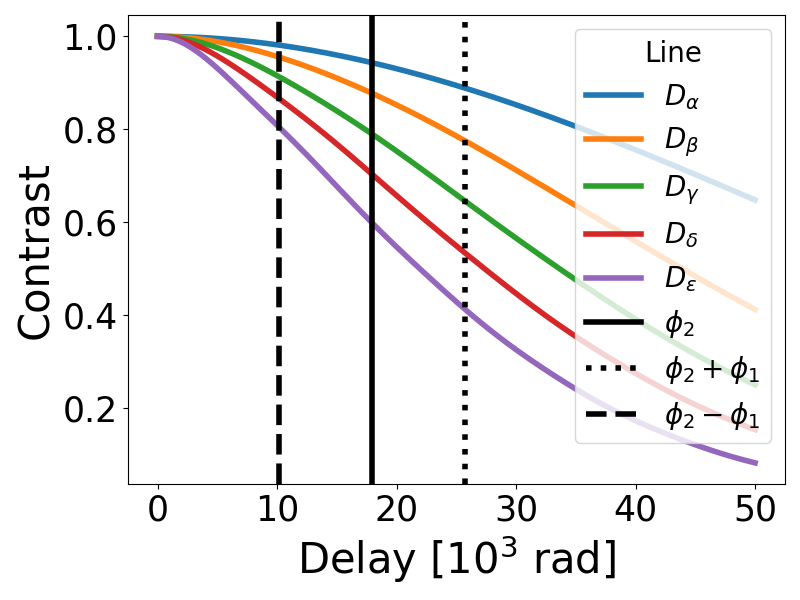}
\caption{}
\end{subfigure}
\vspace{-0.3 cm}\caption{Contrast dependence on delay for a plasma with $n_e = 1e19$ m$^{-3}$, $T_n = 1$ eV and (a) varying levels of hydrogen contamination or (b) different choices of observed spectral line}
\label{fig:fig4_delay_curve_H}
\vspace{-0.3cm} 
\end{figure*}

Zeeman splitting can be accounted for in the strong field approximation using a simple model that depends on the strength of the magnetic field, the angle between the front polarizer, and the projection of the magnetic field on the polarizer surface ($\rho_{pol}$), and the angle between the field and the line of sight connecting the plasma to the camera pixel ($\theta$). The difference in the resulting contrast compared to the more sophisticated Rosato model\cite{rosato_model_2017}, which self-consistently accounts for both Stark broadening and Zeeman splitting, is much smaller than experimental uncertainties for small magnetic fields \cite{Allcock_CIS_density}. For a given lineshape in the absence of a magnetic field $g(\nu)$, the Zeeman-split spectrum will be given by \cite{allcock_thesis}

\begin{align}
    g^Z(\nu) = &   I^{\pi}g(\nu) + I^{\sigma}g(\nu + \nu_B) + I^{\sigma}g(\nu -\nu_B) \\
    I^{\pi} =  & \frac{\sin(\theta)^2}{2}\cos(\rho_{pol})^2 \hspace{1cm} \nu_B =  \frac{e B}{4\pi m_e} \\
    I^{\sigma} = &  \frac{1 + \cos(\theta)^2 -  \sin(\theta)^2\cos(2\rho_{pol})}{4}
\end{align}

The effect of varying levels of background and of Zeeman splitting on the measured contrasts is shown in figure \ref{fig:fig3_delay_curve_background}.  

It can be seen how background emission has the potential to significantly affect the measured contrast and so it is included as an additional free parameter to be fit in the model, along with electron density and neutral temperature. Zeeman splitting, instead, has a minor effect. It mostly affects the $\phi_1 + \phi_2$ term, increasing the slope of the curve, and so will result in a slight overestimation of the temperature and underestimation of the density. A quantitative assessment of neglecting it in the lineshape model is made in \ref{sec:appendix}.

Another contribution that can significantly affect the measured contrast is the presence of impurities. While no impurity lines are expected in the $D_\gamma$ bandpass region, a non-negligible hydrogen concentration can often be found in deuterium plasmas. The presence of the $H_\gamma$ line at 434.04 nm will effectively transform the emission spectrum into a multiplet-like structure, similar to what is observed in CIS impurity studies based on the CIII line. The effect of different hydrogen fractions is shown in figure \ref{fig:fig4_delay_curve_H}.\\  

Due to the multiplet-like structure, the contrast dependence loses its monotonicity on delay, possibly making it a sensitive diagnostic against the presence of impurities or as a means of detecting isotope ratios. In particular, a measured contrast on the difference term ,$\zeta_{2-1}$, lower than the contrast on the sum term,$\zeta_{2}$, will be indicative of the presence of significant hydrogen contamination, as otherwise $\zeta_{2}$ $<$ $\zeta_{2-1}$ would be expected.

The sensitivity of the diagnostic can be tuned more towards Stark or Doppler broadening by choosing an appropriate spectral line. Higher Balmer lines will be more sensitive to Stark broadening, but will also be generally dimmer, reducing the signal-to-noise ratio or requiring longer exposure times. The contrast dependence of the first 5 lines in the Balmer series is shown in figure \ref{fig:fig4_delay_curve_H} for fixed plasma parameters. The MAST-U CIS system has been designed to work with either $D_\gamma$ or $D_\delta$, but only $D_\gamma$ measurements have been assessed in this work as $D_\delta$ measurements can be affected by trace nitrogen impurities \cite{allcock_thesis}.

\section{Tomographic inversions}

While observing an inhomogeneous plasma, such as a tokamak divertor, the normalized spectrum reaching each pixel $j$ will be line integrated along its line of sight and weighted by the local emissivity $\epsilon(\textbf{r})$

\begin{equation}
    \tilde{g}_j(\nu) = \frac{\int_L \epsilon(\textbf{r}) g(\nu, \textbf{r}) \dd \textbf{r}}{\int_L \epsilon(\textbf{r})  \dd \textbf{r}}
\end{equation}

The corresponding three contrast values $\tilde{\zeta}_2, \Tilde{\zeta}_{2+1}$ and $\Tilde{\zeta}_{2-1}$  can be obtained by substituting the spectrum $\tilde{g}(\nu)$ in equation \ref{eq:gamma} and using the three delay values corresponding to the MAST-U instrument. In the assumption of toroidal symmetry, the measured values can be inverted into a 2D profile of density and neutral temperature. As the number of pixels is much larger than the number of cells used to describe the 2D profiles, the problem is overconstrained and it can be solved using a non-linear optimization algorithm. This is similar to the inversion of 2D emissivity profiles usually performed in tokamaks with conventional cameras \cite{carr_inversions_JET}\cite{perek_emissivity_TCV}, with the main difference being the non-linear dependence of the contrast on density and temperature. 

The 2D relative emissivity profile of the measured spectral line is one of the inputs of the analysis and it is assumed known in the synthetic testing. While experimentally it can be obtained with a tomographic inversion of the DC component of the CIS interferogram, in this work the 2D $D_\gamma$ emissivity profile from the Multi-wavelength-imaging diagnostic (MWI) \cite{wijkamp_MWI_2023} is used to ensure consistency with its physics analysis. 

\subsection{Optimization algorithm}

In this work the poloidal profiles are discretized using barycentric linear interpolation, thus the parameters are inferred on the vertices of a triangular mesh \cite{tokamesh}. For a set of vectors describing the discretized poloidal profiles $ \left \{\vb{n_e},\vb{T_n},\vb{C} \right \} $ and a given lineshape model $g(\nu, n_e, T_n, C)$, a matrix with the spectrum of each vertex as a row can be built $G(\nu,\vb{n_e},\vb{T_n},\vb{C})$. The emissivity-weighted line integration can be performed via matrix multiplication with a weighting matrix $W$. It will have elements $W_{jk}=G_{j,k} \epsilon_k$, equal to the elements of the geometry matrix typically used in emissivity tomographic inversions $G$, multiplied by the emissivity of the corresponding vertices $\epsilon_k$. Thus the matrix of line-integrated spectra reaching each pixel $\tilde{G}(\nu,\vb{n_e},\vb{T_n},\vb{C})$ will be
\begin{equation}
    \tilde{G}(\nu,\vb{n_e},\vb{T_n},\vb{C}) = WG(\nu,\vb{n_e},\vb{T_n},\vb{C}) \label{eq:spectra_mat}
\end{equation} 

The three contrast values, indexed by $q$, can then be obtained by numerical integration over the frequency, indexed by $l$
\begin{equation}
    \tilde{\zeta}_{q,j} = \lvert \sum_l \tilde{G}_{j,l}\exp \left\{ i \phi_{j,q} \left[1 + \kappa_0 \left(\frac{\nu_l-\nu_0}{\nu_0} \right) \right]\right\} \Delta \nu \rvert
\end{equation} 
The optimization can then be cast as a likelihood maximization problem
\begin{equation}
    \mathcal{L}^\zeta(\vb{n_e}, \vb{T_n}, \vb{C}) = - \sum_{q,j} \frac{\lvert \zeta^\star_{q,j} - \tilde{\zeta}_{q,j}(\vb{n_e}, \vb{T_n}, \vb{C})\rvert^2}{\sigma_{q,j}^2} \label{eq:likelihood}
\end{equation}
with $\zeta^\star_{q,j} \in \left[ \zeta^\star_{2,j}, \zeta^\star_{2+1,j}, \zeta^\star_{2-1,j} \right]$ the contrast values measured on pixel j and $\sigma_{q,j}$ their uncertainties. 

If accounted for, the Zeeman splitting can be included by substituting equation \ref{eq:spectra_mat} with
\begin{equation}
    \tilde{G}^Z(\nu) = W^\pi G(\nu) +  W^\sigma G(\nu + \nu_B) +  W^\sigma G(\nu - \nu_B)
\end{equation} 
where $W^\pi_{j,k} = I^\pi_jW_{j,k} $, $W^\sigma_{j,k} = I^\sigma_jW_{j,k} $ and the dependencies on the inferred vectors are implied.

Smoothing penalties must be included in the optimization to avoid overfitting the data and inferring unrealistically discontinuous profiles. This is done by penalizing the difference between the value on each vertex and the average of the values on all of its connected vertices. Using the "umbrella" operator $U$ \cite{tokamesh}, this can be expressed as a matrix multiplication with the corresponding parameter vector. Taking for example the smoothing constraint for the density 
\begin{equation}
     \mathcal{L}^S_{n_e} = - S_{n_e}\sum_{j,k} U_{j,k}n_e^k
\end{equation} 
where $S_{n_e}$ is a parameter setting the strength of the smoothing penalty for the density parameter. 

The total likelihood to be maximized will then be 
\begin{equation}
     \mathcal{L} = \mathcal{L}^\zeta + \mathcal{L}^S_{n_e} + \mathcal{L}^S_{T_n} + \mathcal{L}^S_{C}  
\end{equation}

If the delay shear across the detector is negligible, $\delta \phi(x,y) << \phi_D$, the optimization can be greatly simplified. In this case, the numerical integration has no pixel dependence and it can be performed for each cell before the line integration if Doppler shifts are assumed small \cite{Howard_Overview_CIS}

\begin{equation}
    \tilde{\zeta}(\vb{n_e},\vb{T_n},\vb{C}) = W \zeta(\vb{n_e},\vb{T_n},\vb{C})  
\end{equation} 

The variation of $\phi_2$ across the entire image is negligible, but $\phi_{2+1}$ and $\phi_{2-1}$ vary of $\pm 2\%$ across the sensor. This is small enough to still give meaningful results and can be used if fast optimizations are required, but large enough to justify including the delay shear in the model. A quantitative assessment of neglecting the delay shear is given in \ref{sec:appendix}.  In this work, all minimizations have been performed with the L-BFGS optimizer.

\subsection{Uncertainty estimation}\label{sec:MC}

By running the inversion multiple times with different initial conditions and perturbations on the input data, the uncertainty in the inferred parameters can be estimated in a Monte Carlo approach. The inversions routine is run multiple times with perturbed inputs and the resulting 2D profiles are used to build a probability distribution for the parameters of each cell. For each sample, it includes:
\begin{itemize}
    \item \textit{Noise in the contrast measurements}. Uncorrelated Gaussian noise is added to the contrasts of each pixel, with standard deviations representative of experimental conditions.  The same $\sigma_{q,j}$ in equation \ref{eq:likelihood} are used to generate the noise samples.
    \item \textit{Contrast calibration uncertainty}. The CIS contrast measurements are calibrated by measuring instrumental contrast images with a spectral lamp, analogously to measuring the instrumental broadening in a spectrometer, and dividing the measured contrast images by the calibration ones \cite{Doyle_CIS}. A systematic uncertainty with a uniform distribution in [-3\%, 3\%] is assumed on each of the calibration images.
    \item \textit{Uncertainty in the 2D emissivity profiles}. A multivariate normal distribution is used to model the uncertainty in the normalized emissivity profiles. The correlation between different pixels is assumed to scale inversely with their distance, while the noise strength is assumed to be 0.1\% of the maximum emissivity in the profile. This is a typical uncertainty based on a Monte Carlo analysis of the emissivity tomographic inversions. The elements of the correlation matrix are given by
    \begin{align}
       & \sigma^E_{i,j} \propto \frac{1}{\abs{d_{ij}}^{0.1}} & \sigma^E_{i,i} = 1\cdot 10^{-3} E_{max} & 
    \end{align}
    By including some correlation between nearby vertices, the inclusion of strong noise-like patterns in an otherwise smooth input profile is avoided.
    \item \textit{Uncertainty in the electron temperature}. The Stark broadening profile has a small dependence on the electron temperature (equation \ref{eq:stark_profile}). The electron temperature is assumed the same for all vertices and sampled from a log-uniform distribution in the range [0.5, 5] eV. 
    \item \textit{Uncertainty in the front polarizer angle $\rho_{pol}$}. For non-longitudinal views, Zeeman splitting will be dependent on $\rho_{pol}$. This can significantly affect the results in the presence of high fields, but due to the low magnetic field in the MAST-U divertor ($\leq$ 0.7 T) it mostly acts as a correction. As the angle of the front polarizer and the polarization state of the light at the front polarizer are not known, a uniform distribution of $\rho_{pol} \in [0, \pi/2]$ is assumed.
\end{itemize}

The inferred 2D profiles for each sample are used as the inputs of a Gaussian kernel density estimate of the probability distribution for the parameters of each vertex. The maximum of the distribution is used as the estimate for the parameter, while the highest density interval surrounding the maximum and containing 68 \% of the distribution is used as the confidence interval and thus a metric for the uncertainty.

\subsection{Broadening model uncertainty}

\begin{figure}[h!]\centering
\includegraphics[width= 0.49\textwidth]{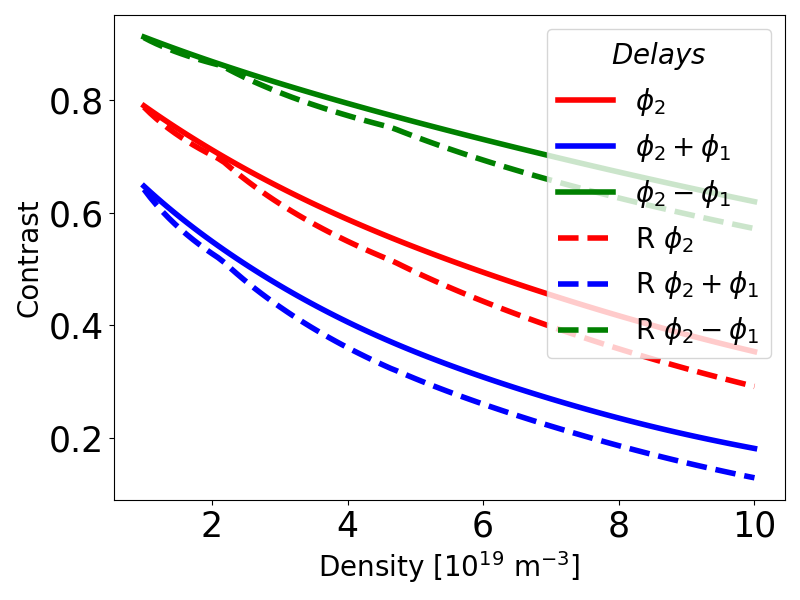}
\vspace{-0.6 cm}\caption{Contrasts modelled for a 1 eV homogeneous plasma using the modified Lorentzian fit ($\phi_2,\phi_{2+1},\phi_{2-1}$) and corresponding values modelled using Rosato's lineshape model (R $\phi_2$,R $\phi_{2+1}$,R $\phi_{2-1}$)}
\label{fig:fig5_rosato_comp}
\vspace{-0.3cm} 
\end{figure}

An incorrect Stark broadening model will be reflected in incorrect $n_e$ and $T_n$ profiles. To evaluate the uncertainty in the broadening model, it can be compared to the lineshape modeled by Rosato et al \cite{rosato_model_2017}. The three contrasts corresponding to the two broadening models as a function of density for a 1 eV plasma are compared in figure \ref{fig:fig5_rosato_comp}.
It can be seen how the modified Lorentzian fit systematically overestimates all three contrasts compared to the other model. This would correspond to a $\sim$ 15 \% overestimation of the density for typical experimental conditions in the MAST-U divertor. While the difference is relatively small and is below the uncertainty in the inferred densities, this systematic error can still be partially accounted for with a correction factor on the measured contrast images. 

After inverting a set of contrast images, the inferred 2D profiles can be used to forward model three contrast images with the Rosato model. These will have systematically lower contrasts than the input images and can be used to determine a set of correction factors, defined as the relative error between the forward-modeled contrast images with the two models. It is then possible to run the inversions again after increasing the contrasts of the input images by the correction factors. The resulting 2D profiles will lead to forward-modeled contrast images that do not underestimate the measured contrasts when using the Rosato model. The difference between the original and corrected 2D profiles can be compared to the uncertainties from the Monte Carlo and is another source of uncertainty in the inversion.    

\section{Synthetic testing} \label{sec:Synthetic_test}

\begin{figure*}[h!]\centering
\includegraphics[width= 0.95\textwidth]{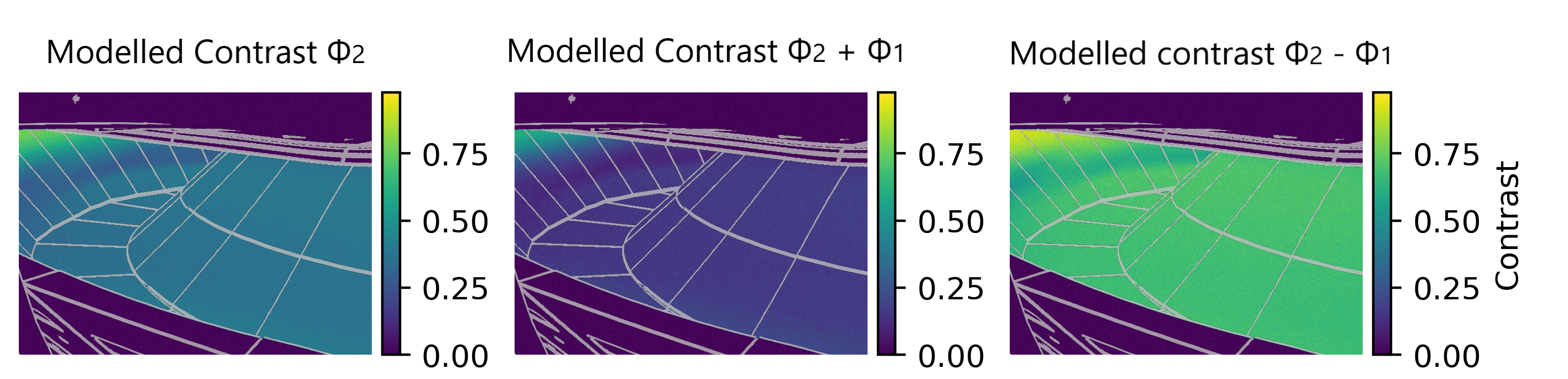}
\vspace{-0.0 cm}\caption{Forward modelled $\phi_2,\phi_{2+1}$ and $\phi_{2-1}$ contrast images based on 2D profiles from a Super-X SOLPS simulation.}
\label{fig:fig6:SOLPS_contrast_images}
\vspace{-0.3cm} 
\end{figure*}
\begin{figure}[h!]\centering
\includegraphics[width= 0.49\textwidth]{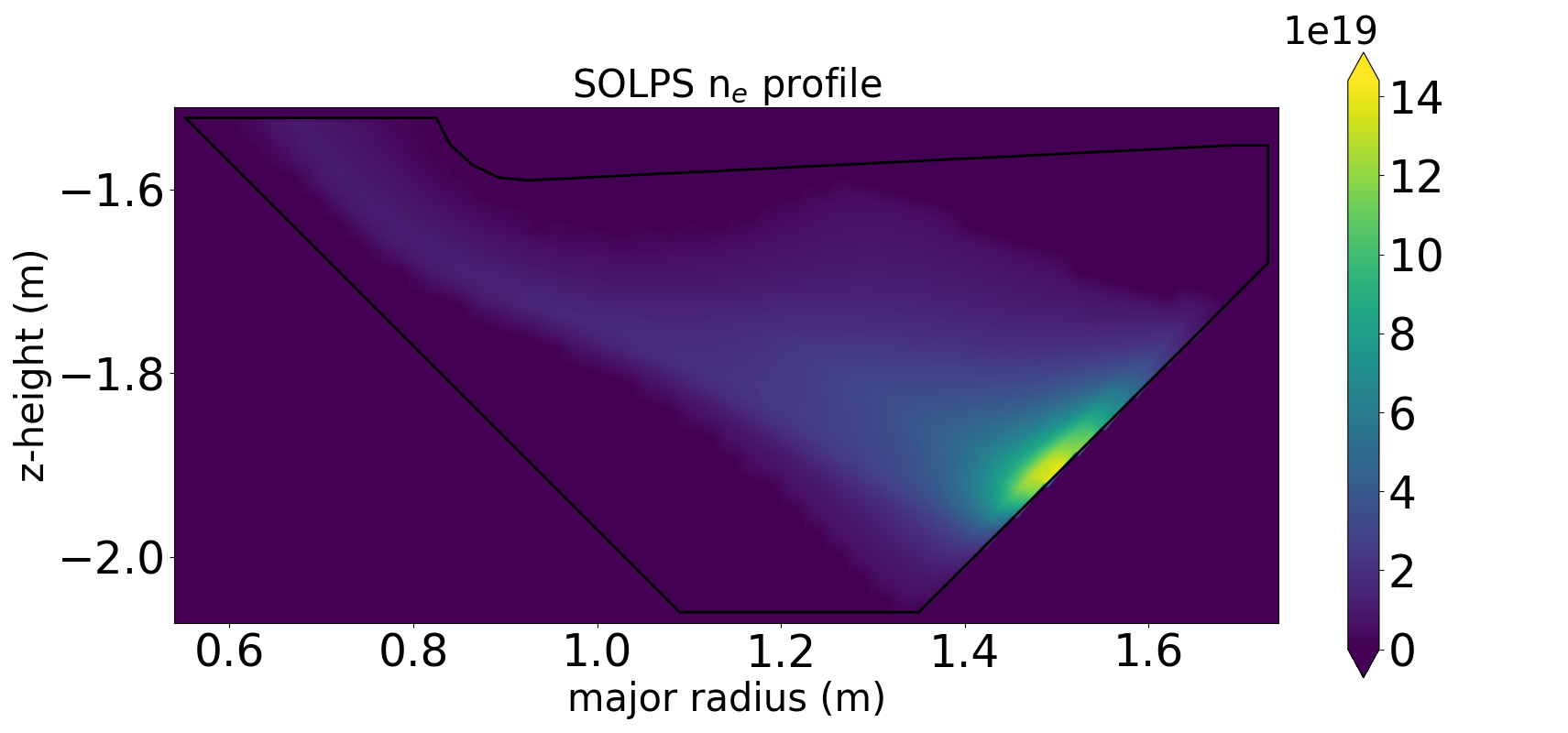}
\includegraphics[width= 0.49\textwidth]{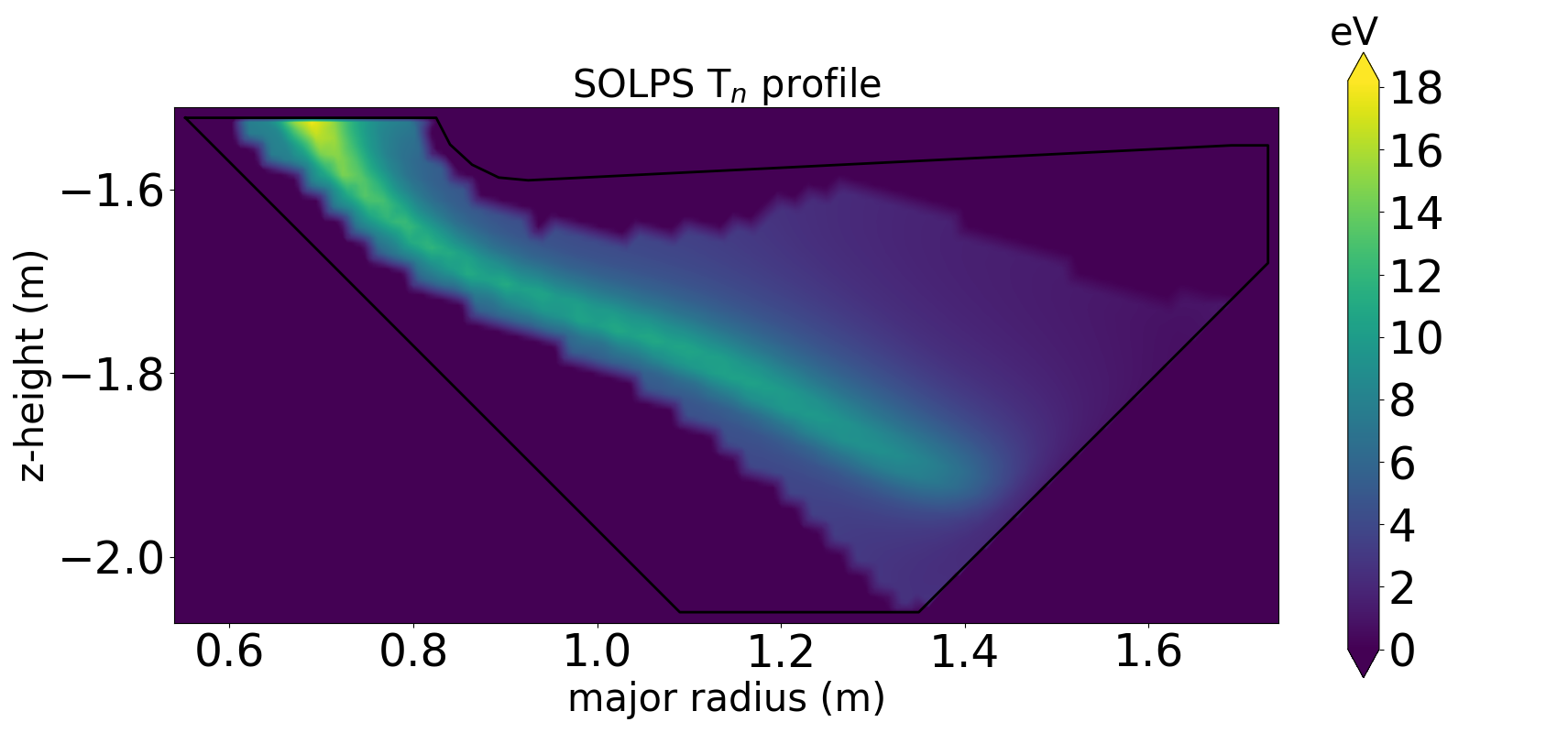}
\includegraphics[width= 0.49\textwidth]{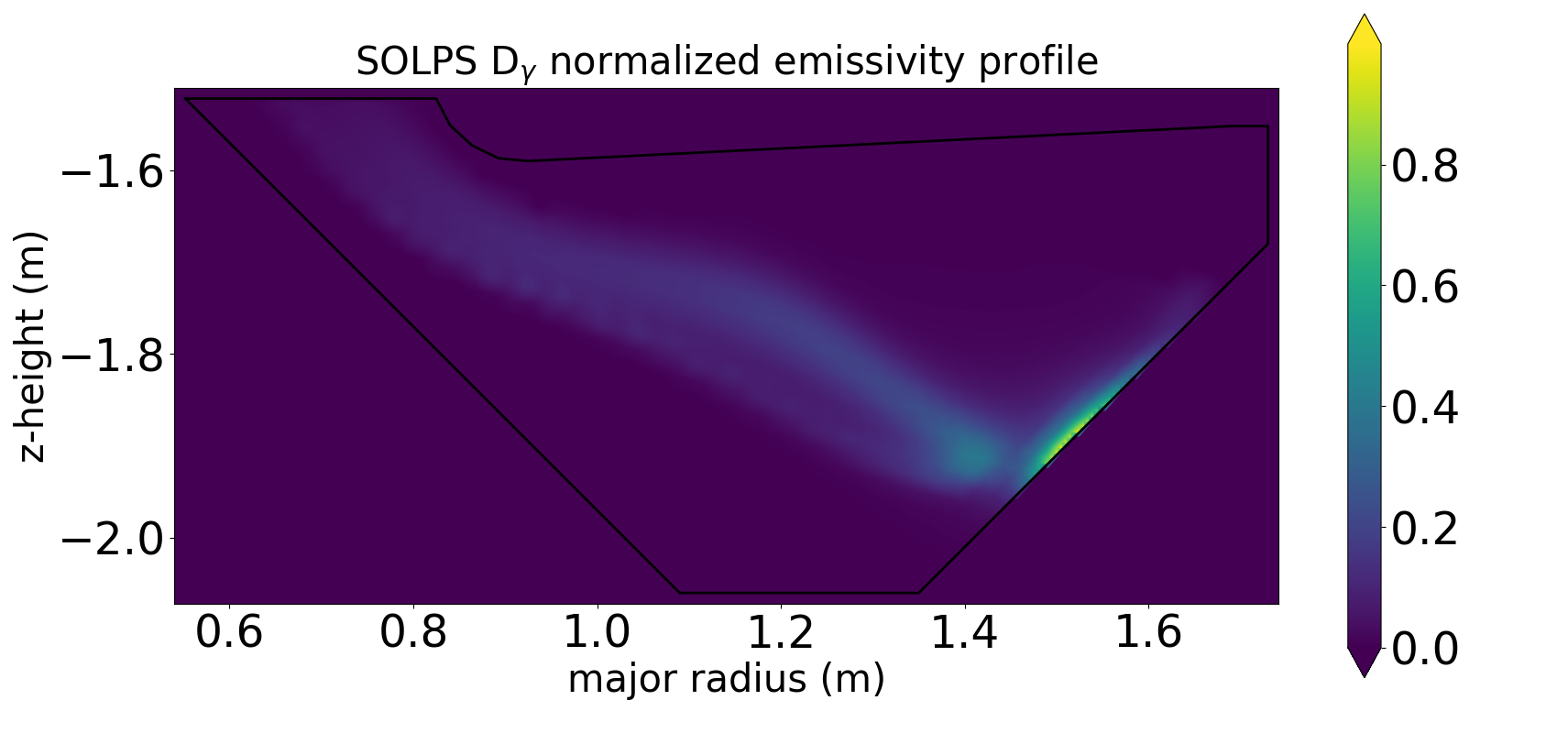}
\vspace{-0.6 cm}\caption{SOLPS simulation of 2D poloidal profiles of $n_e$, $T_n$ and $D_\gamma$ emissivity in the MAST-U Super-X divertor}
\label{fig:fig7:SOLPS_inputs}
\vspace{-0.3cm} 
\end{figure}

The performance of the inversion can be characterized by testing it on synthetic contrast images based on profiles from SOLPS-ITER simulations, where the correct solution of the inversion is known. The 2D $n_e$, $T_n$, and $D_\gamma$ emissivity profiles for a SOLPS simulation of the MAST-U Super-X divertor with 2.5 MW of input power, a 4 $\cross 10^{21} s^{-1} $ deuterium fuelling rate, and no impurity seeding are shown in figure \ref{fig:fig7:SOLPS_inputs}.
\begin{figure}[h!]\centering
\includegraphics[width= 0.49\textwidth]{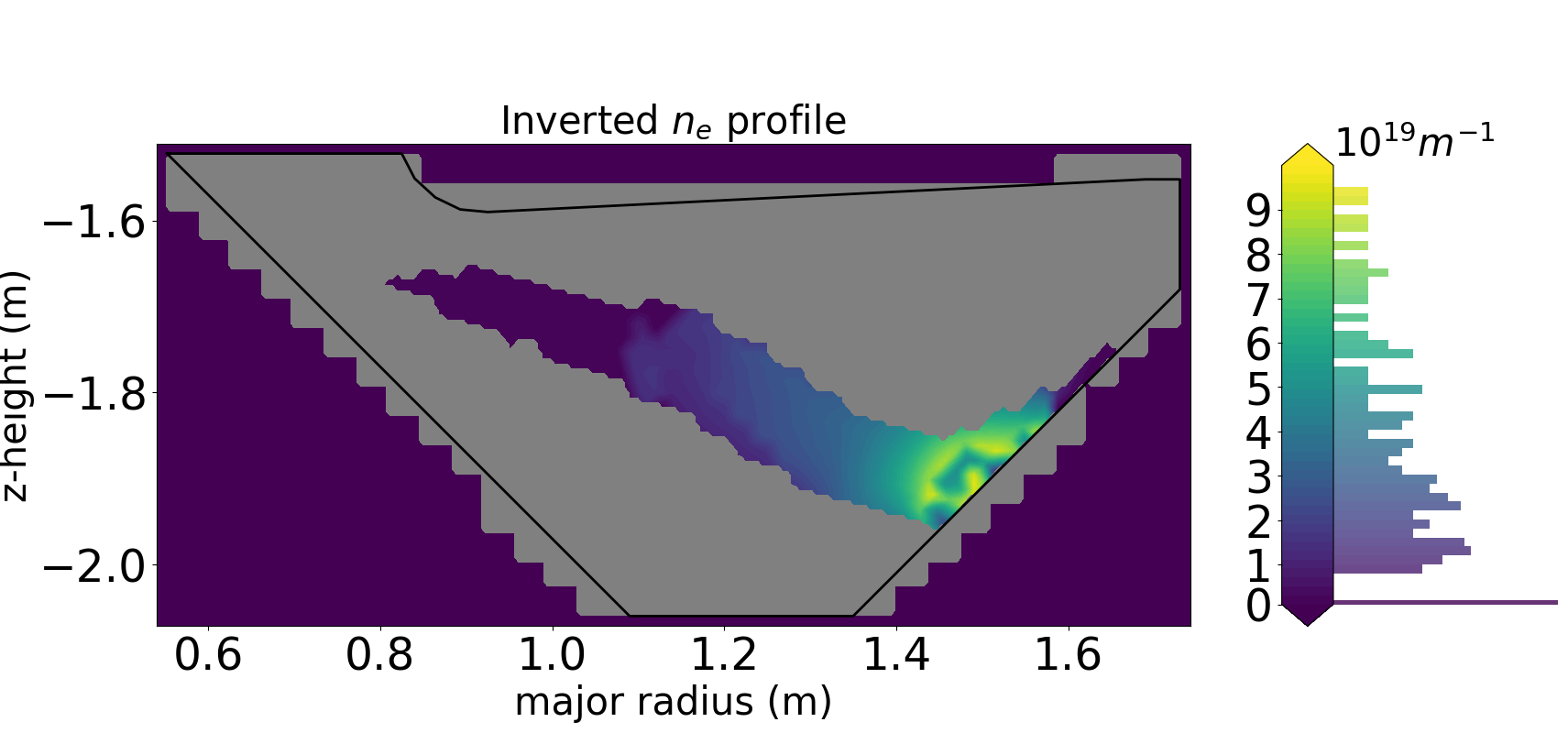}
\includegraphics[width= 0.49\textwidth]{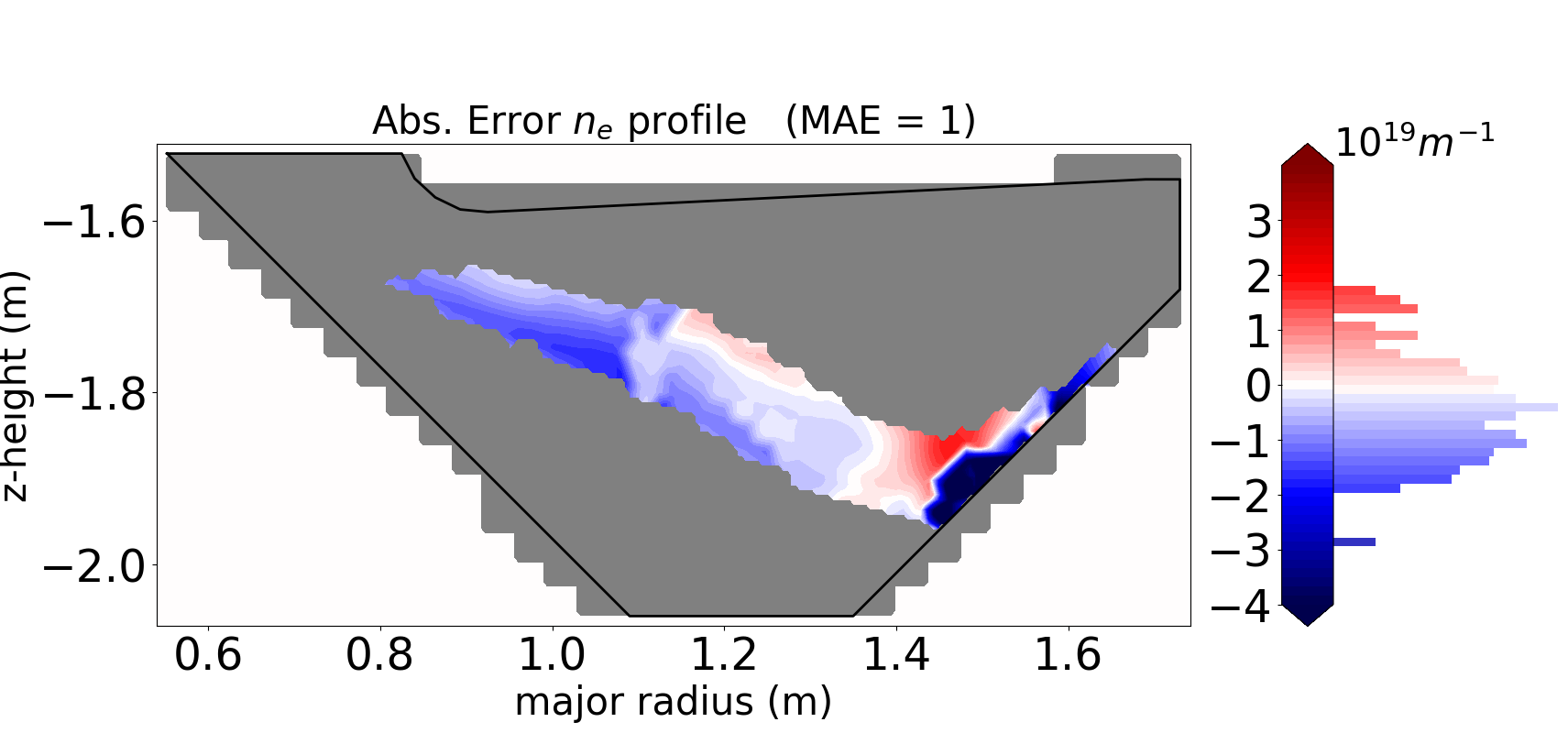}
\vspace{-0.0 cm}\caption{Inferred electron density profile from the synthetic contrast images and error compared to original profile. The error is masked in the region with emission above 5 \% of the maximum emissivity and the mean absolute error is of 1 $\cdot 10^{19} m^{-3}$.}
\label{fig:fig8_synth_ne}
\vspace{-0.3cm} 
\end{figure}
\begin{figure}[h!]\centering
\includegraphics[width= 0.49\textwidth]{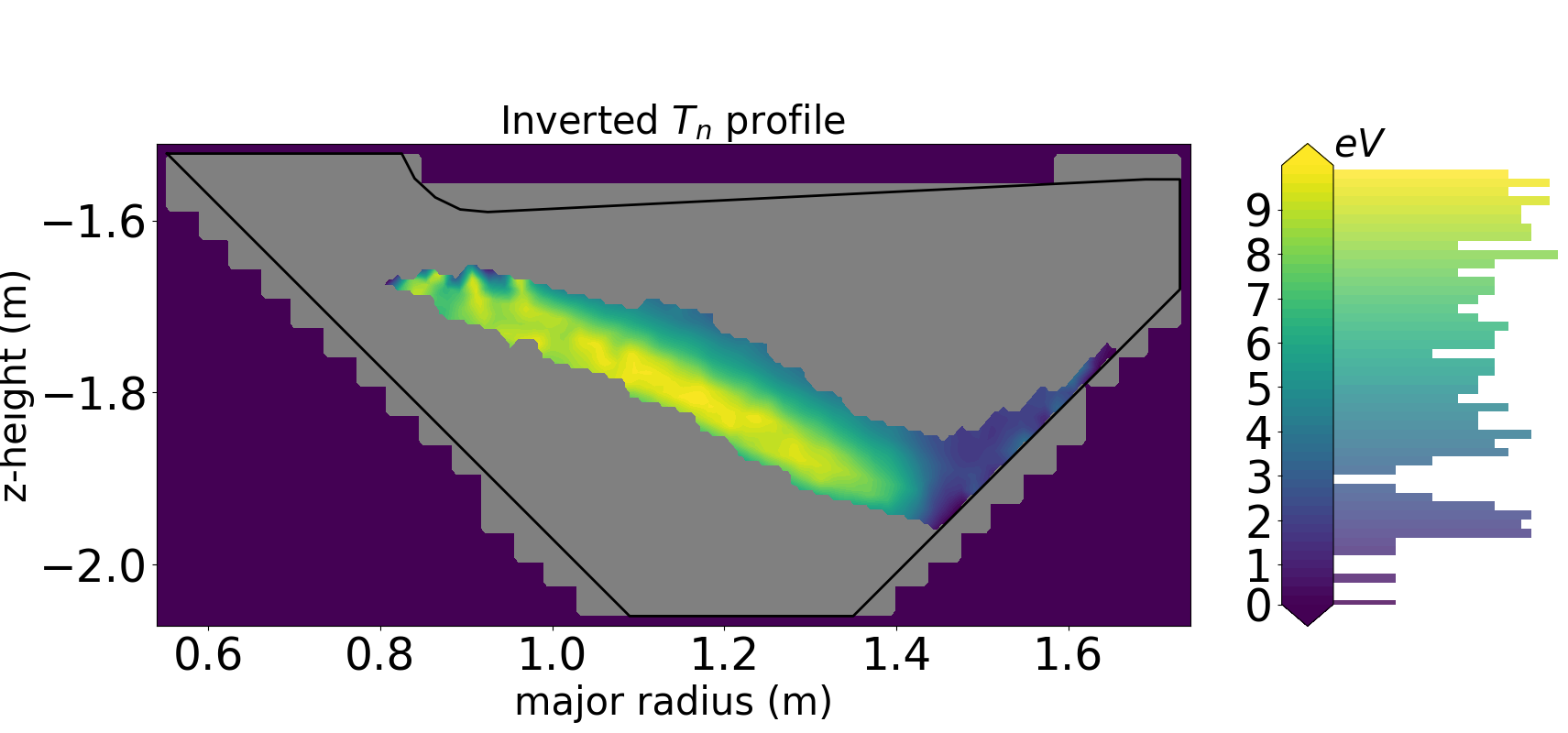}
\includegraphics[width= 0.49\textwidth]{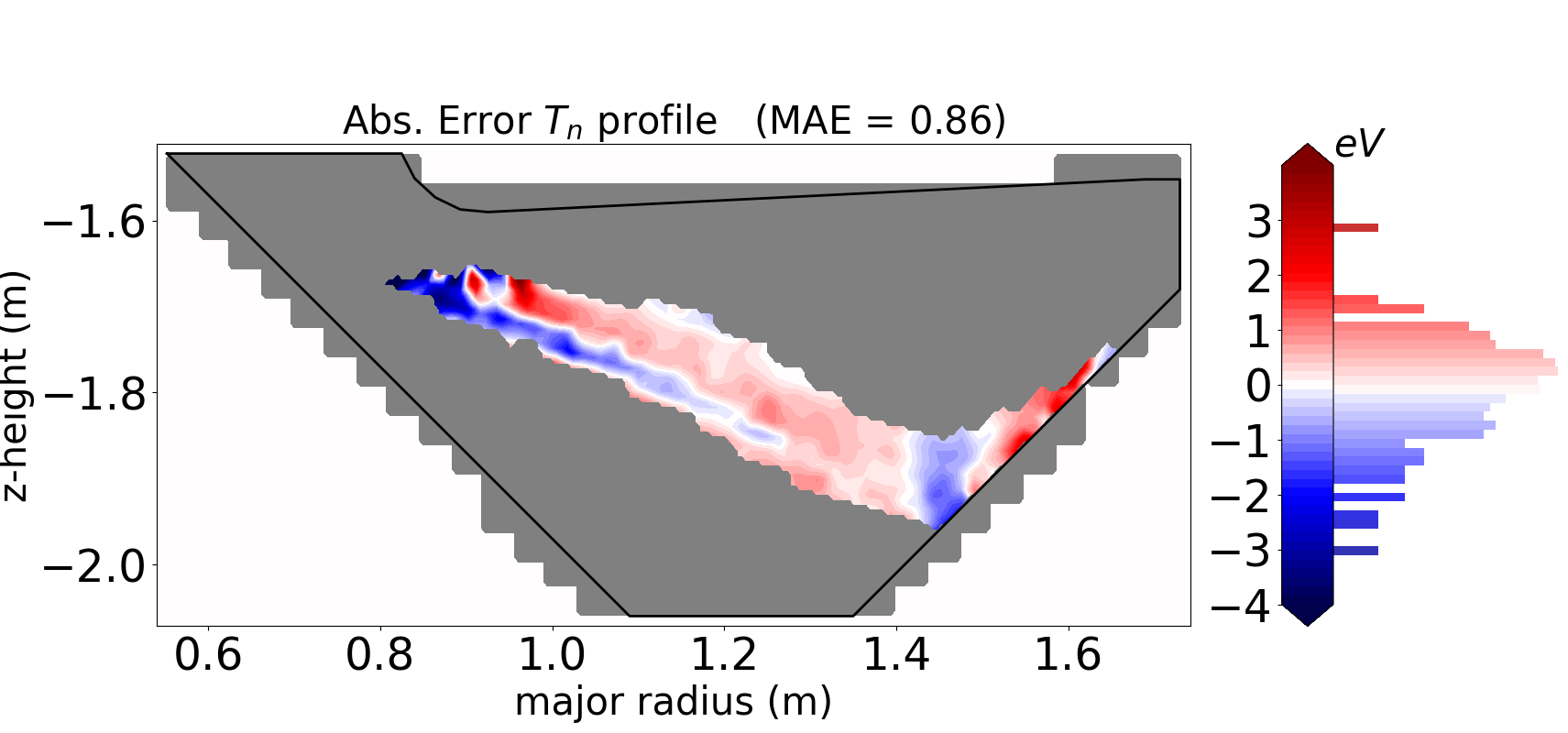}
\vspace{-0.0 cm}\caption{Inferred neutral temperature profile from the synthetic contrast images and relative error compared to original profile. The error is masked in the region with emission above 5 \% of the maximum emissivity and the mean absolute error is of 0.86 eV.}
\label{fig:fig9_synth_Tn}
\vspace{-0.3cm} 
\end{figure}
The 2D emissivity profile is modeled using the PEC coefficients from ADAS, assuming $n_i = n_e$, and not accounting for molecular effects
\begin{align}
        \epsilon = n_0n_ePEC_{ex}(T_e, n_e) + n_e^2PEC_{rec}(T_e, n_e)
\end{align}

The corresponding synthetic contrast images generated using the modified Lorentzian model  are shown in figure \ref{fig:fig6:SOLPS_contrast_images}.

The modeled background emission is assumed to be the same at all wavelengths in the width of the bandpass filter and to be coming purely from bremsstrahlung. Its strength is written in terms of a user set parameter $C_0$ and the expected dependencies on $n_e$ and $ T_e$
\begin{align}
        C = C_0\frac{n_e^2}{\sqrt{T_e}} 
\end{align}
Only the regions of the contrast images where the line-integrated emissivity is larger than 15 \% of the maximum are included in the inversion, to emulate experimental conditions where some regions of the image are too dark to provide contrast measurements with low signal-to-noise ratio. The emissivity profile used as input in the inversion is the same used to generate the synthetic images, with the addition of the Monte Carlo correlated noise described in section \ref{sec:MC}.
Gaussian noise with $\sigma_{q,j} = 0.025$ for all three contrasts is included in the Monte Carlo, and the same value is assumed as uncertainty in the inversion. This is comparable to the demodulation uncertainty measured during the calibrations \cite{Doyle_CIS}. 
The inferred density profile and its error  are shown in figure \ref{fig:fig8_synth_ne}.
Analogously, the inferred $T_n$ profile and its error are shown in figure \ref{fig:fig9_synth_Tn}.
\begin{figure}[h!]\centering
\includegraphics[width= 0.49\textwidth]{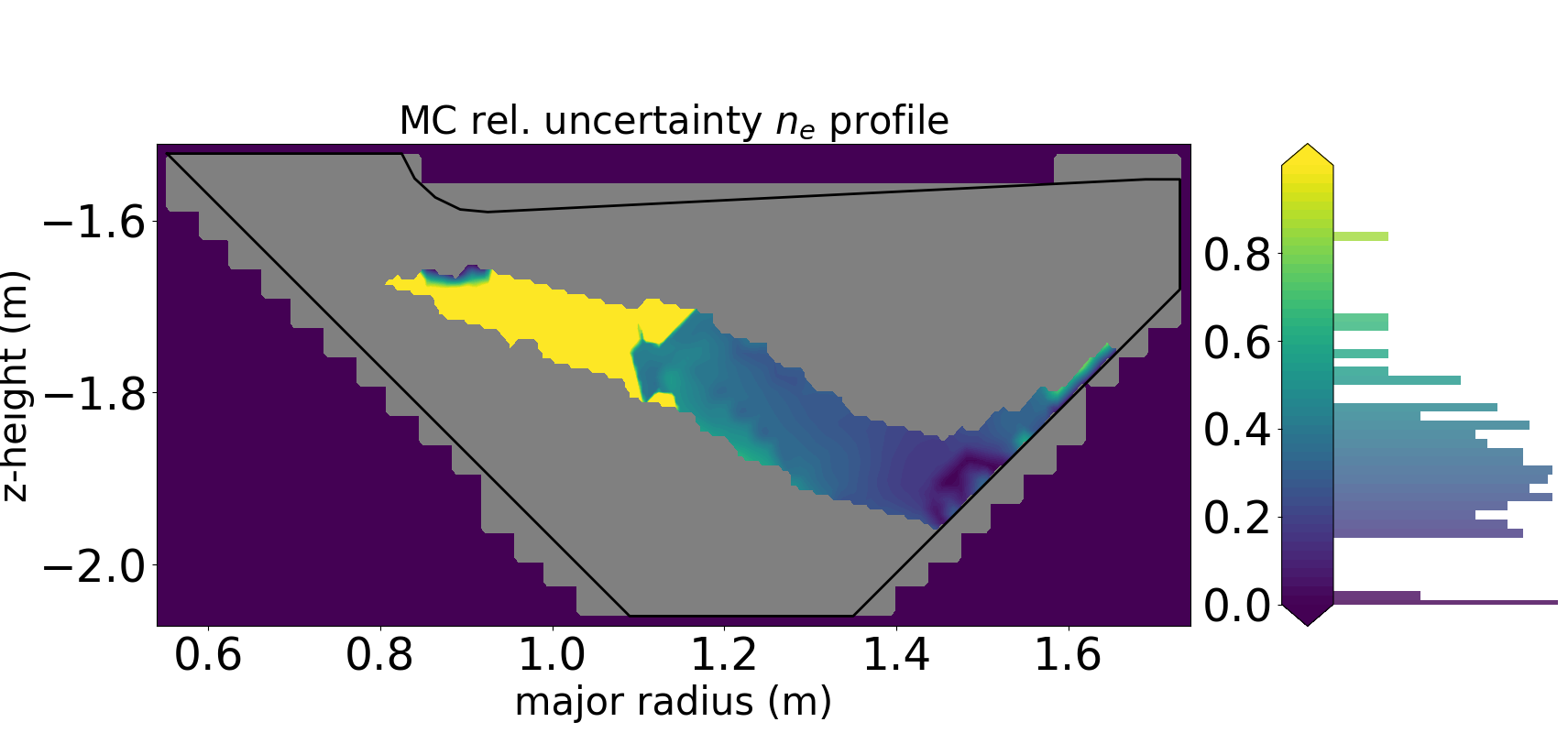}
\vspace{-0.0 cm}\caption{Inferred relative uncertainty in the density profile using the Monte Carlo analysis Masked-off region with emissivity above 5 \% of the maximum emissivity.}
\label{fig:fig10_synth_uncert}
\vspace{-0.3cm} 
\end{figure}
\begin{figure}[h!]\centering
\includegraphics[width= 0.49\textwidth]{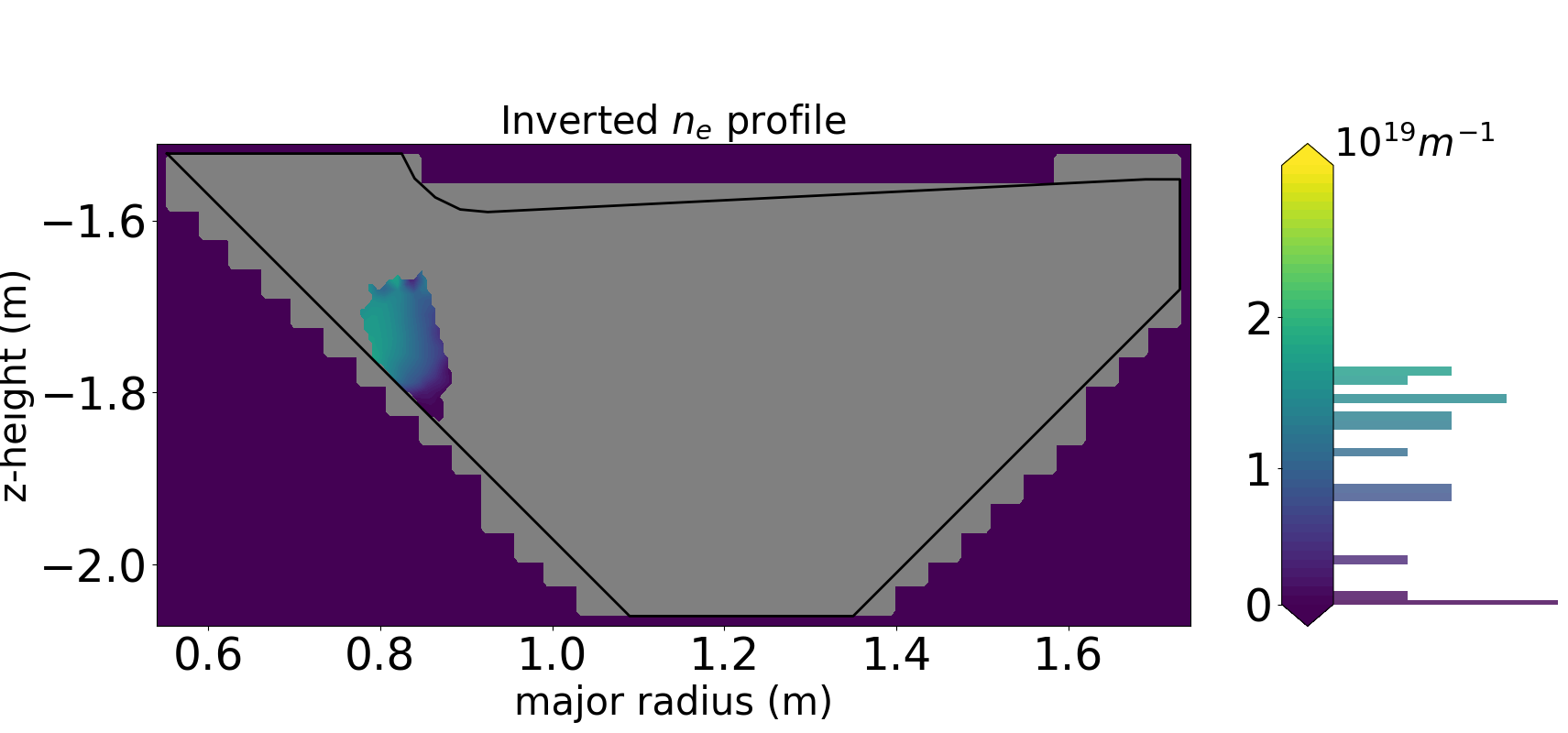}
\includegraphics[width= 0.49\textwidth]{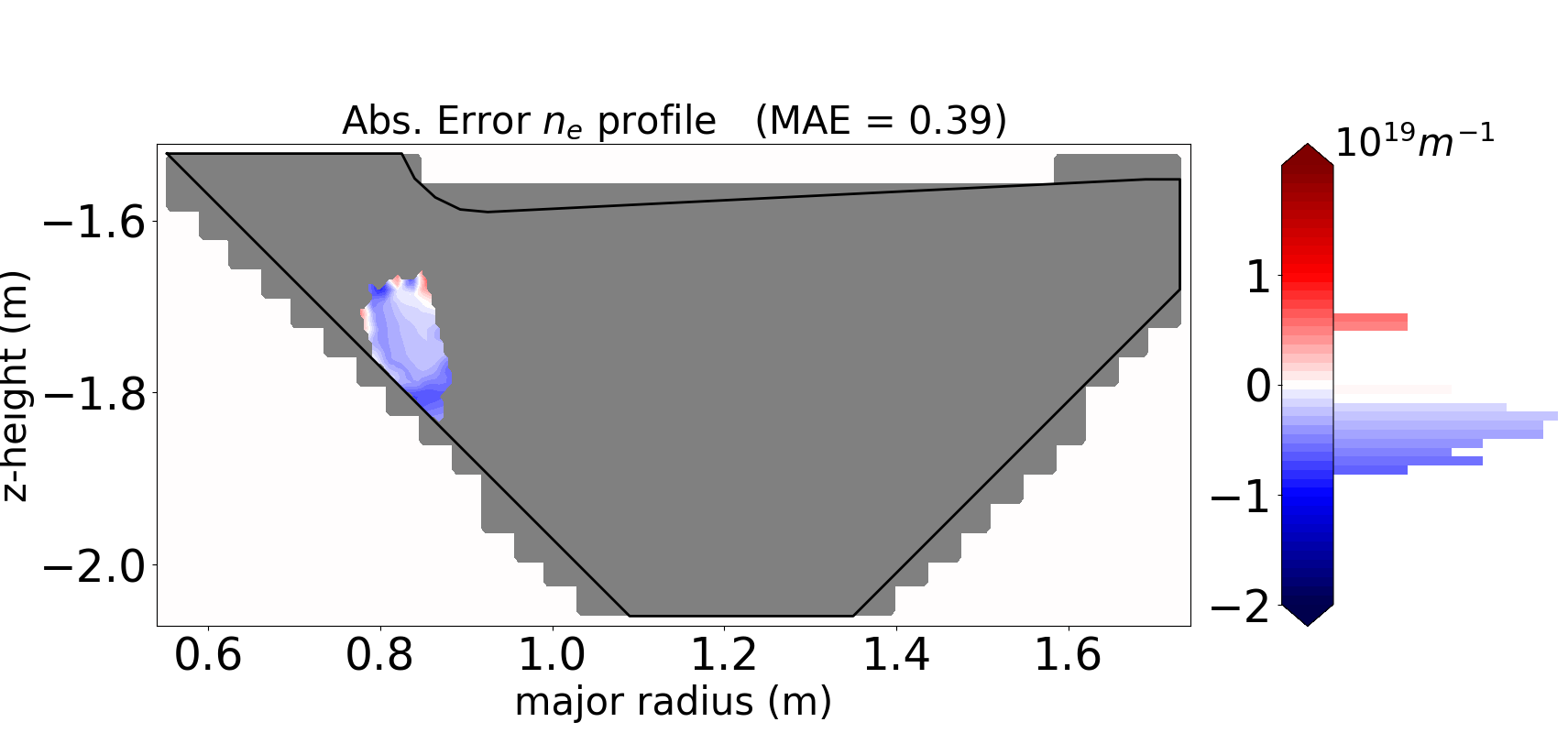}
\vspace{-0.0 cm}\caption{Inferred density profile and error for the synthetic case of a conventional divertor. Masked off region with emissivity above 5 \% of the maximum emissivity and the mean absolute error is of 0.39 $\cdot 10^{19} m^{-3}$.}
\label{fig:fig11_synth_CD}
\vspace{-0.3cm} 
\end{figure}
It can be seen how the error distribution for the density profile, plotted along the colorbar, has a mean absolute value (MAE) of 1 $\cdot 10^{19} m^{-3}$ and it is mostly contained in the [-2 $\cdot 10^{19} m^{-3}$,+ 1 $\cdot 10^{19} m^{-3}$] range across a spatial profile with strongly varying density. This MAE is comparable in magnitude with typical uncertainties on line-averaged electron density measurements obtained with spectroscopy in comparable conditions \cite{Verhaegh_thesis}, but have the advantage of providing a 2D profile instead of line-averages. A higher error is visible very close to the target, probably due to a lack of tangential sightlines. The mean absolute error for the $T_n$ profile is of 0.86 eV, while the error distribution is mostly contained in the [-1.5 eV, +1.5 eV] range.  

The mean absolute relative error in the masked region is of 42\% and 25 \% for the density and temperature profiles respectively. In both cases the distribution is mostly contained in the [-20 \%, + 20 \%] range, but the average value is brought up by the regions of high error very close to the target, where the relative error can reach 60 \%, and towards the top of the image, where the density is low and thus the relative error high. By comparison, the median absolute relative errors in the masked region are less affected by small regions of high errors and are of 29 \% and 9.3 \% respectively.  An anti-correlation between the density and temperature errors can be noticed, with regions where one parameter is overestimated corresponding to the underestimation of the other. This can be immediately understood as a correct inference of the broadening of the line but an insufficient resolution to distinguish perfectly between the two broadening mechanisms.

The estimated relative uncertainty obtained from 30 Monte Carlo samples is shown in figure \ref{fig:fig10_synth_uncert}. 
\begin{figure*}[]\centering
\includegraphics[width= 0.95\textwidth, trim={2.5cm 1cm 0.75cm 0.75cm},clip]{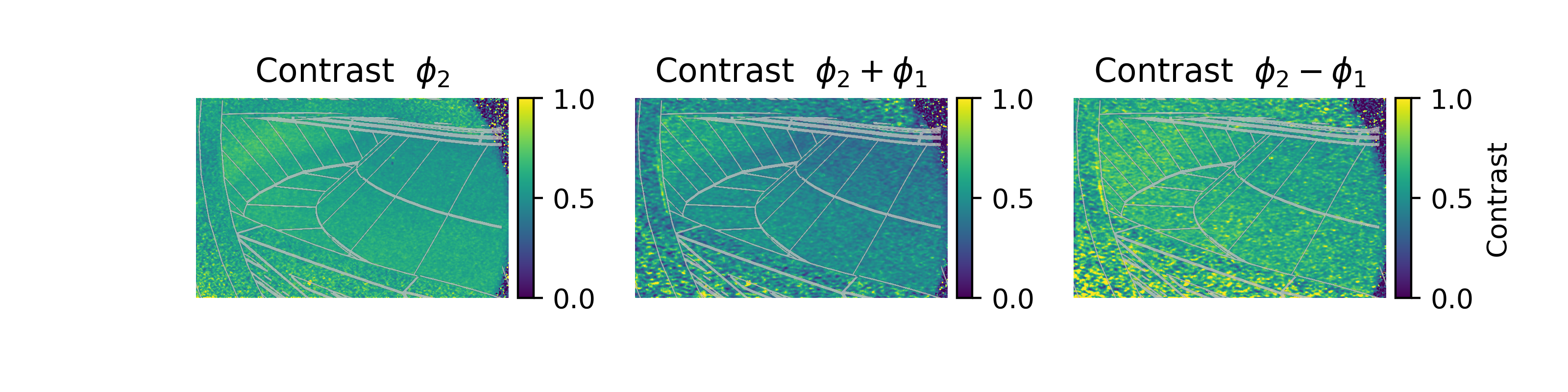}
\vspace{-0.3 cm}\caption{Calibrated contrast images at t = 0.66 s in shot \#46893. The wireframe of the MAST-U CAD model is overlaid on the image.}
\label{fig:fig13_exp_cont}
\vspace{-0.3cm} 
\end{figure*}
Typical estimated relative uncertainties are in the range [15 \%, 45 \%], with lower estimated uncertainties in regions of stronger emission. A comparison between the inferred uncertainty and the inference error is given in \ref{sec:appendix_error}.
Comparable performance in relative error is obtained when testing the analysis on the SOLPS simulation of a conventional divertor with $ 1.2 \cross 10^{20} s^{-1}$ deuterium fuelling rate, as shown for the density profiles in figure \ref{fig:fig11_synth_CD}.   
In this case, the strike point is at the entrance of the divertor chamber, limiting the volume where significant light is emitted. The mean absolute relative error of 39 \% across the profile is comparable to the Super-X case and suggests that even though the view is not optimized for the conventional divertor, meaningful comparisons between the two divertor configurations can still be made. 

\section{Application to experimental data}
\begin{figure}[]\centering
\includegraphics[width= 0.49\textwidth]{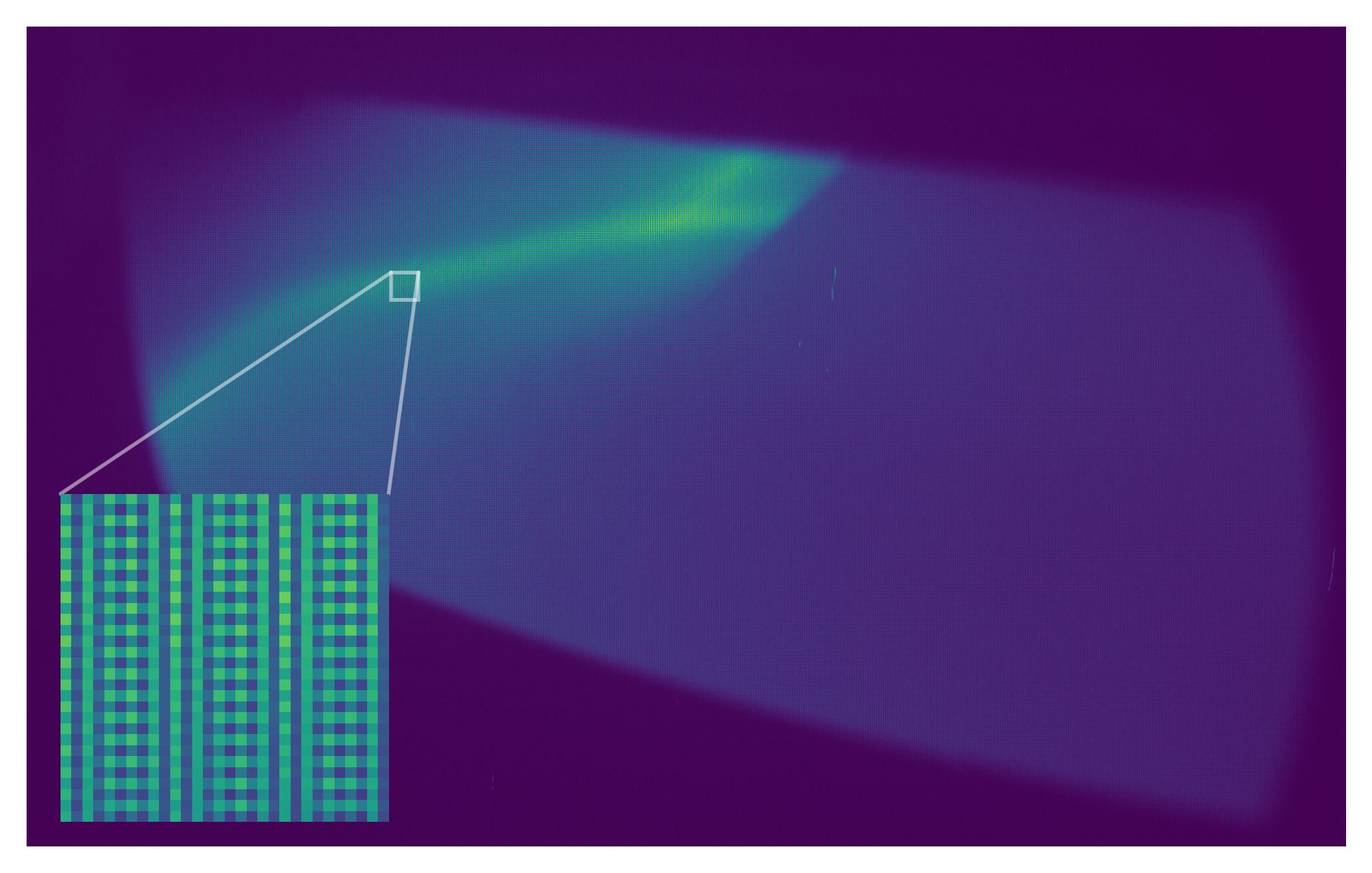}
\vspace{-0.3 cm}\caption{Raw interferogram measured by the CIS instrument at t = 0.66 s in shot \#46893.}
\label{fig:fig12_exp_interf}
\vspace{-0.3cm} 
\end{figure}

The analysis has been tested on experimental measurements taken during a core fuelling scan in a beam-heated L-mode scenario with a double-null Super-X divertor configuration, shot \#46893. The discharge has a plasma current of 750 kA, 1.75 MW of off-axis NBI power and the fuelling scan has been performed from 25 \% to 50 \% of the Greenwald density using a low-field side valve to inhibit H mode access. The divertor is detached throughout the entire Super-X phase of the discharge. The raw interferogram and corresponding demodulated contrast images at t = 0.66 s, corresponding to a core density of 3.35 $\cross 10^{19}$ m$^{-3}$ (40 \% Greemwald fraction),  are shown in figures \ref{fig:fig12_exp_interf} and \ref{fig:fig13_exp_cont} respectively. 

The resolution of the image is $\sim$ 4 times larger than the MWI images, due to the inability to bin together the pixels on a polarized sensor. To increase the speed of the inversion, the contrast images are downsampled by a factor 2 in each dimension, leading to contrast images of comparable resolution to the MWI. The increased resolution is also used to estimate the noise in the contrast measurements as the standard deviation in a square 9-pixel window centered on each pixel, before the downsampling. A hydrogen fraction of 5\% is assumed, as it minimizes the mean error in the residuals while being in agreement with the spectra measured by the DMS spectrometer. 
\begin{figure}[h!]\centering
\includegraphics[width= 0.49\textwidth]{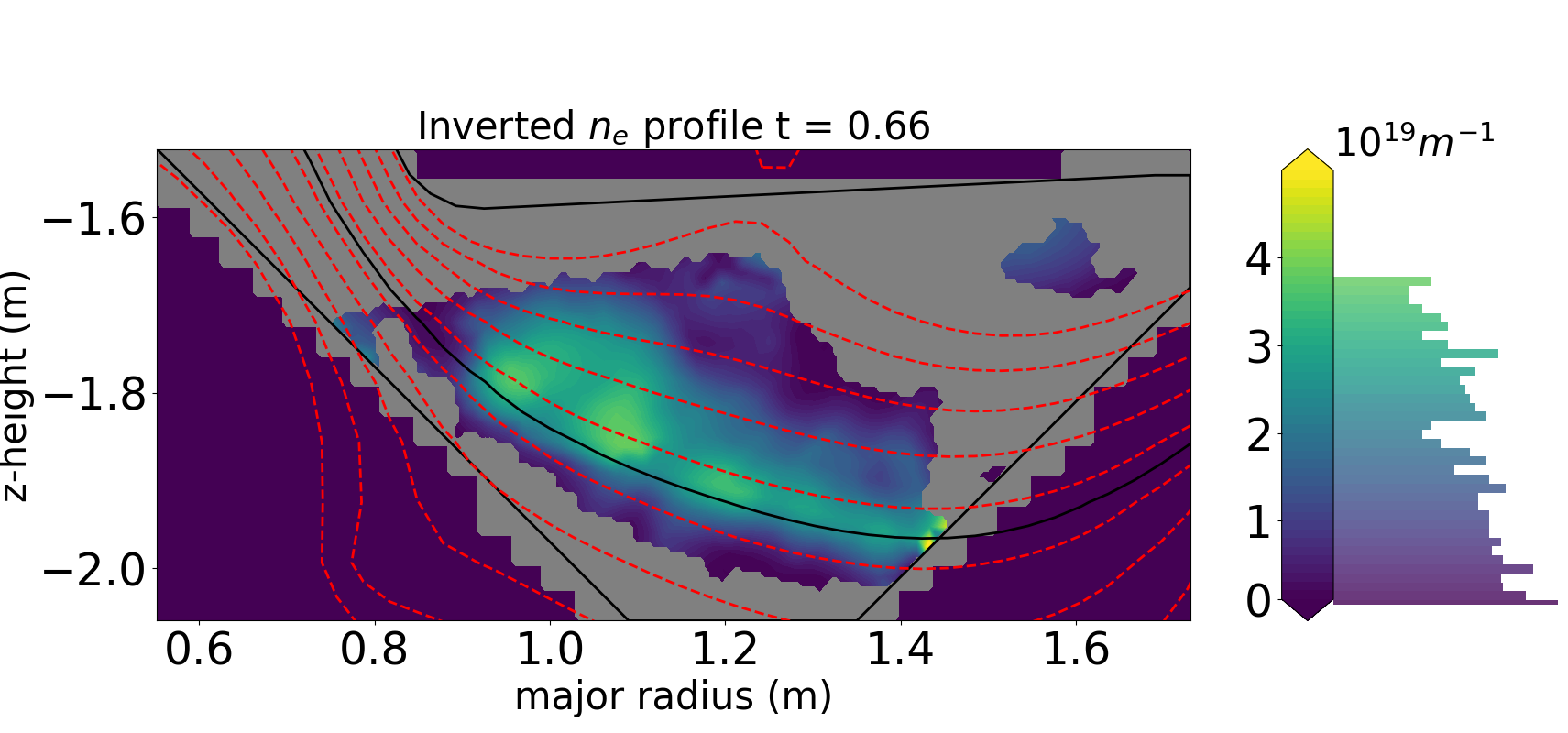}
\includegraphics[width= 0.49\textwidth]{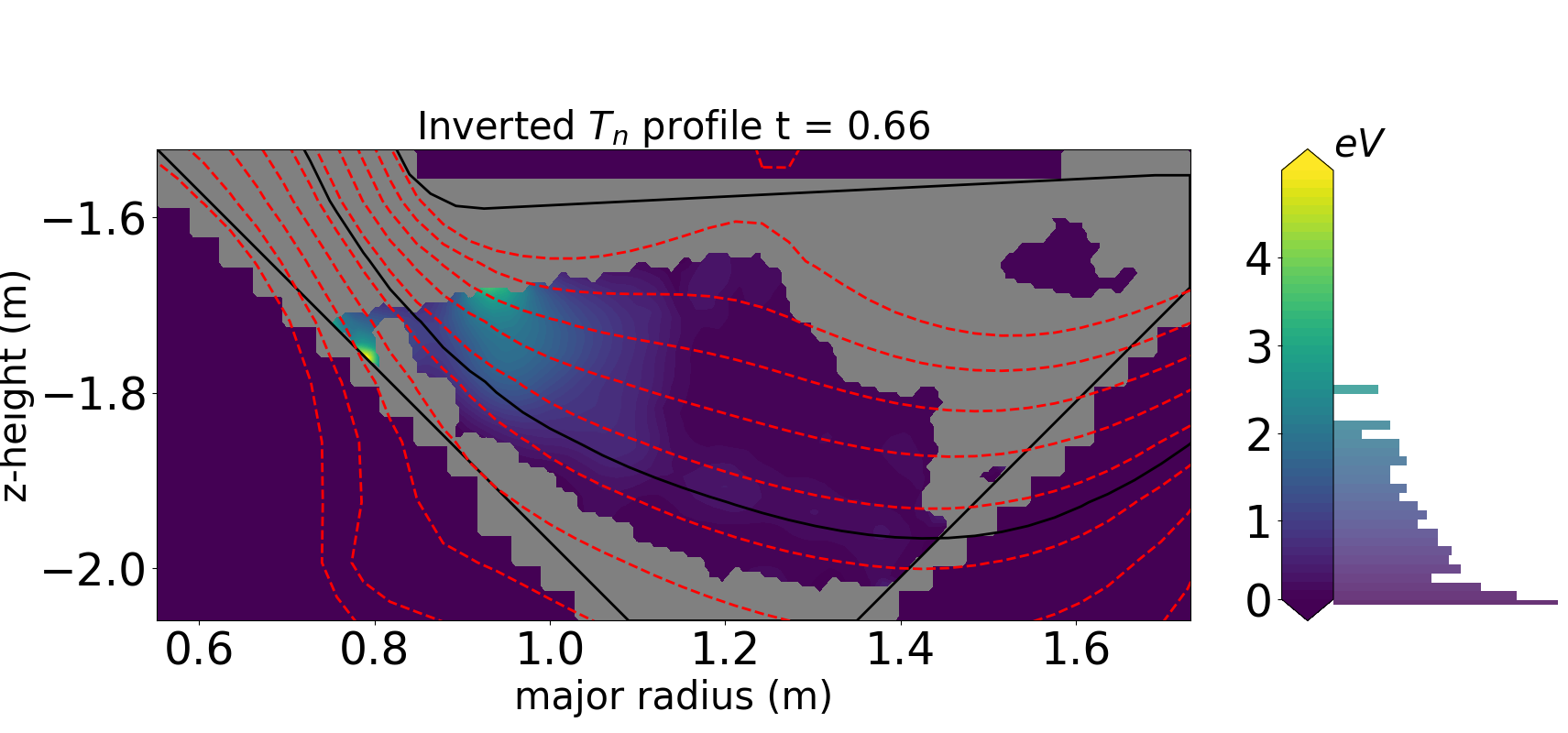}
\vspace{-0.0 cm}\caption{Inferred electron density and neutral temperature profiles at t=0.66 s in shot \#46893. The red lines represent the flux surfaces as reconstructed by EFIT, with the separatrix is shown in black.}
\label{fig:fig14_exp_profiles}
\vspace{-0.3cm} 
\end{figure}\begin{figure*}[]\centering
\includegraphics[width= 0.95\textwidth]{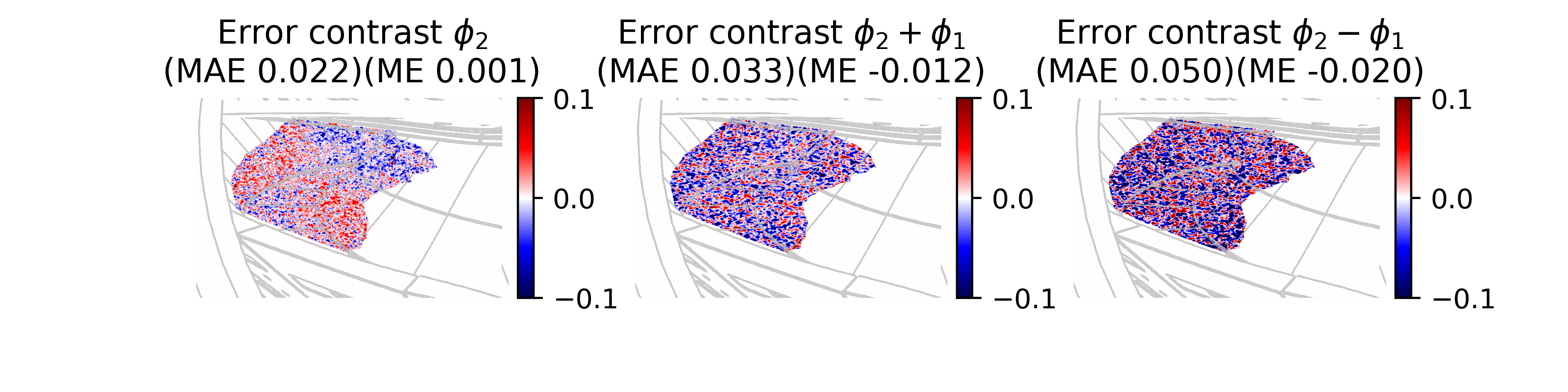}
\vspace{-0.3 cm}\caption{Residuals of the three contrast images for the inversion shown in figure \ref{fig:fig14_exp_profiles}. The mean absolute error (MAE) and mean error (ME) is reported for each image independently.}
\label{fig:fig16_exp_residuals}
\vspace{-0.3cm} 
\end{figure*}
The inferred electron density and neutral temperature profiles from a Monte Carlo inversion with 30 samples are shown in figure \ref{fig:fig14_exp_profiles}.  
Typical electron densities of 2-4 $\cdot 10 ^{19}$ m $^{-3}$ and neutral temperatures of 0-3 eV are observed near the separatrix. The electron density profile does not show strong density gradients along the magnetic field lines, while the neutral atom temperature can be observed to be decreasing from around 3 eV at the top edge of the MWI view to below 1 eV at the target.

\begin{figure}[h!]\centering
\includegraphics[width= 0.49\textwidth]{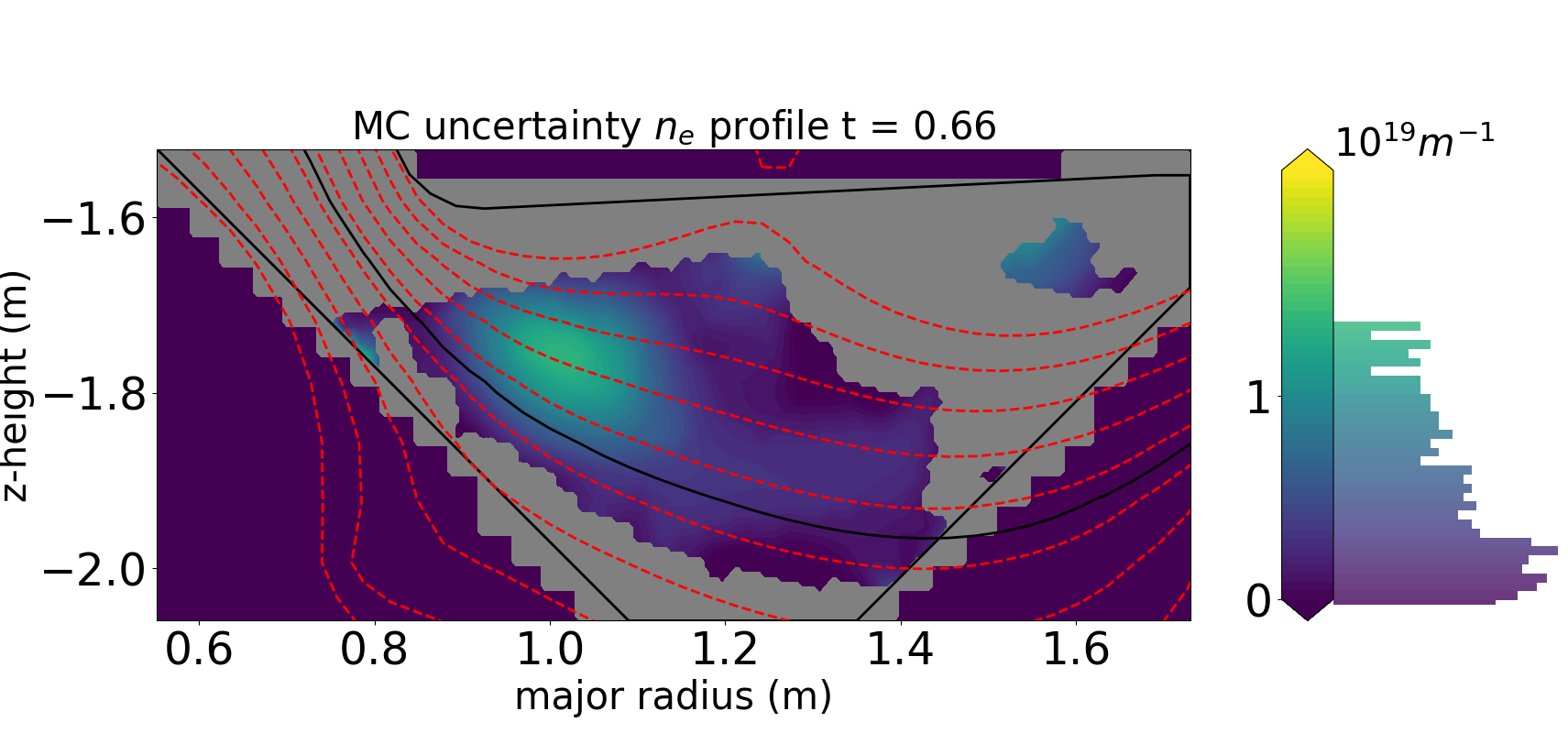}
\includegraphics[width= 0.49\textwidth]{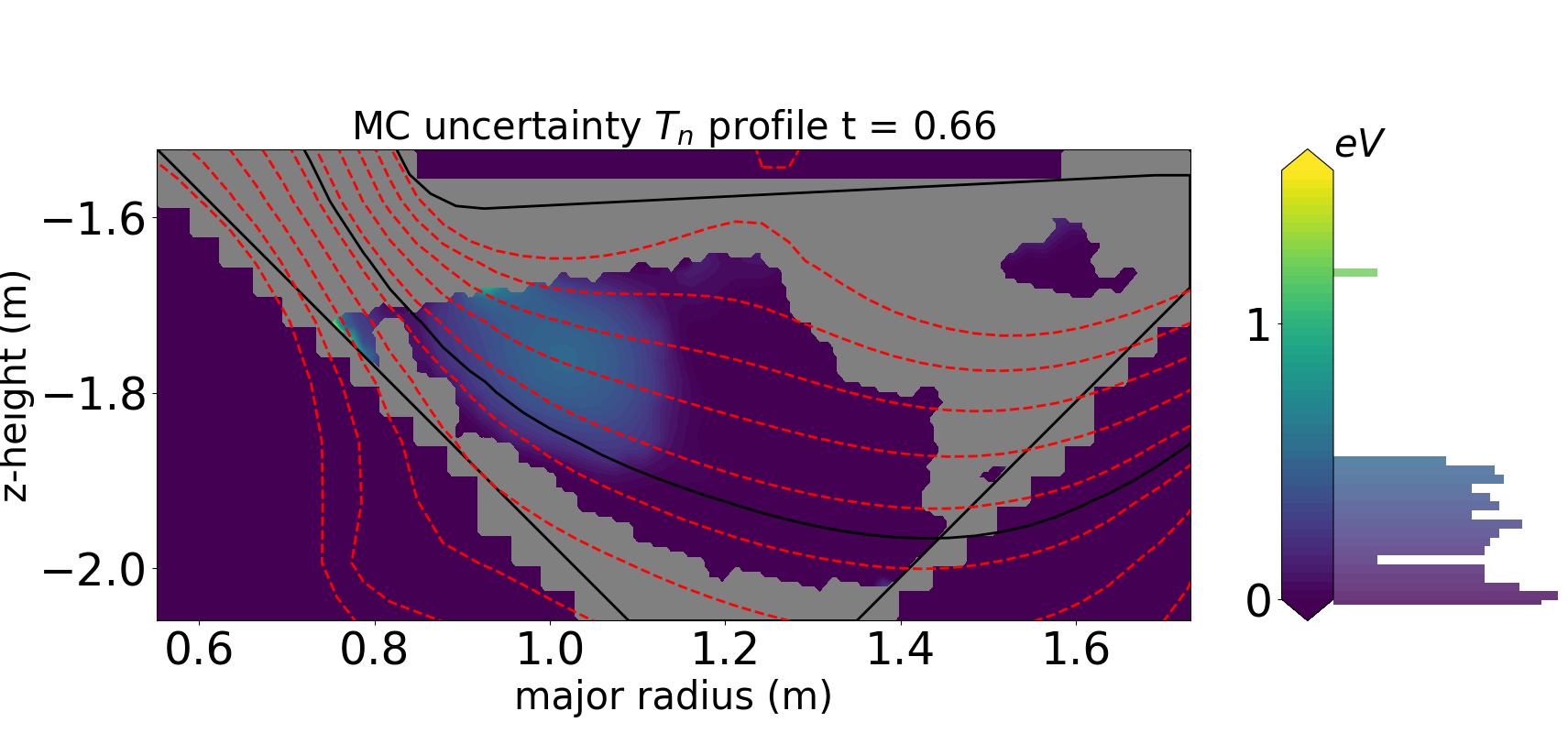}
\vspace{-0.0 cm}\caption{Inferred uncertainties in the electron density and neutral temperature profiles at t=0.66 s in shot \#46893. The red lines represent the flux surfaces as reconstructed by EFIT, with the separatrix shown in black.}
\label{fig:fig15_exp_uncert}
\vspace{-0.3cm} 
\end{figure}

The corresponding relative uncertainties are shown in figure \ref{fig:fig15_exp_uncert}.
The distribution of uncertainties is broad, with higher uncertanties at lower major radius where the emission is lower.
Images of the contrast residuals, the difference between the measured contrast values and the forward modeled images from the results of the inversion, are shown in figure \ref{fig:fig16_exp_residuals}. The residuals show a noise-like behavior, which is the desired outcome if the lineshape model is representative of the data and the deviation from it is mostly due to noise in the measurements. 

The mean absolute errors are comparable with the respective uncertainties in the contrasts determined by the local standard deviation and no clear spatial structure is visible. The residuals of the $\phi_2$ term have zero average, while the $\phi_2+\phi_1$ and $\phi_2-\phi_1$ have a small negative bias in the mean error (ME) of $\sim$ 0.025. While this is considered acceptable as it is below the experimental uncertainty, it may be indicative of a lower hydrogen fraction than the 5\% assumed, as that would leave the $\phi_2$ term mostly unaffected, as shown in figure \ref{fig:fig4_delay_curve_H}.
\subsection{Comparison with other diagnostics}
\begin{figure}[h!]\centering
\begin{subfigure}[b]{0.23\textwidth}
\includegraphics[width= \textwidth]{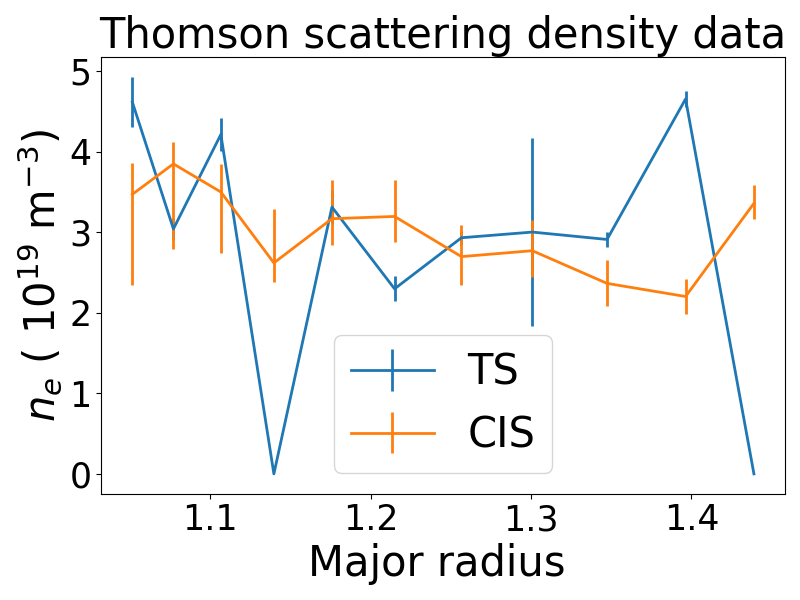}
\caption{}
\end{subfigure}
\begin{subfigure}[b]{0.23\textwidth}
\includegraphics[width=\textwidth]{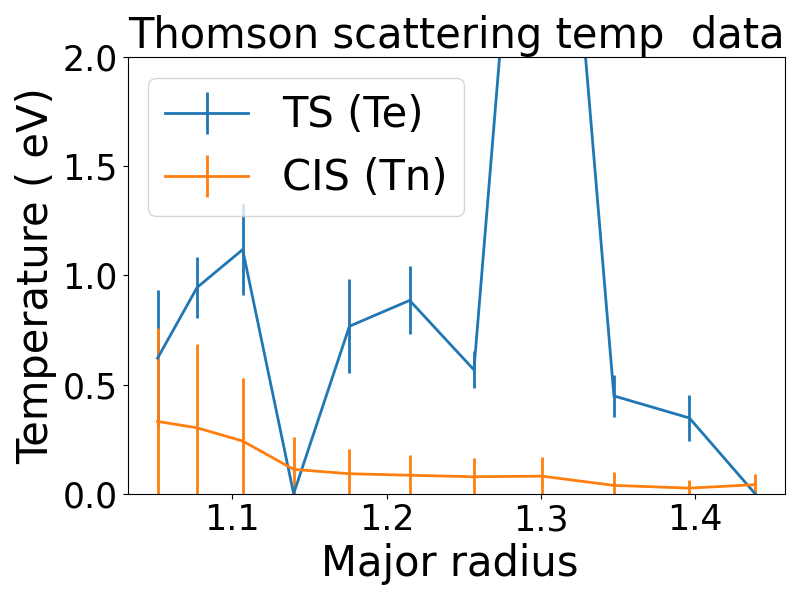}
\caption{}
\end{subfigure}
\vspace{-0.3 cm}\caption{Comparison between the electron density and neutral temperature inferred by the CIS with the electron density and temperature inferred by the Divertor Thomson scattering. }
\label{fig:fig17_TS}
\vspace{-0.3cm} 
\end{figure}

The results of the inversion can be compared to the electron densities and temperatures measured by the divertor Thomson scattering system. The CIS 2D profiles have been interpolated onto the measurement points of the Thomson scattering and a comparison between the two is given in figure \ref{fig:fig17_TS}.
The inferred electron densities are in reasonable agreement, showing a mostly flat spatial profile up to a major radius of R $\sim 1.35$ m. The Thomson scattering shows an increase in density towards the target which is outside the uncertainty inferred by the CIS inversion. This could be due to the low reliability of the inversion technique near the target, or to the low electron temperatures in those two points ($\leq 0.5 $ eV) which can make the Thomson scattering data unreliable. While they are not expected to be the same, both electron temperature and neutral temperature appear to be low ($T_e$ $\leq$ 1.5 eV, $T_n$ $\leq$ 0.5 eV), with the latter consistently lower than the former in the region observed by the Thomson scattering. 

\begin{figure}[h!]\centering
\begin{subfigure}[b]{0.49\textwidth}
\includegraphics[width= \textwidth]{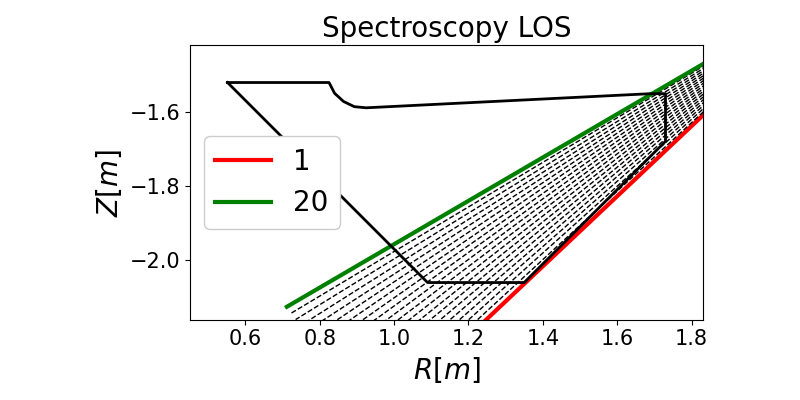}
\caption{}
\end{subfigure}
\begin{subfigure}[b]{0.35\textwidth}
\includegraphics[width=\textwidth]{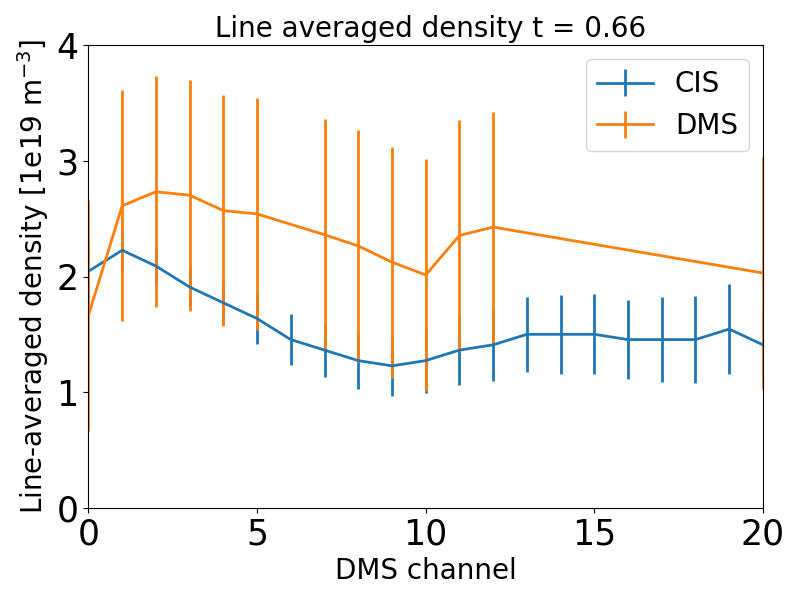}
\caption{}
\end{subfigure}
\vspace{-0.3 cm}\caption{(a) 30 Lines of sight of the MAST-U spectroscopy diagnostics of interest for the comparison. (b) Comparison of the emissivity weighted line averaged electron density inferred by the CIS and by the spectrometers measuring high n Balmer lines.  }
\label{fig:fig18_DMS}
\vspace{-0.3cm} 
\end{figure}

The inferred density profile can also be compared with the emissivity-weighted line-averaged densities estimated by fitting the lineshapes profiles of the high n Balmer lines measured by the divertor spectrometer. While high n Balmer line measurements were not available during shot 46893, the CIS results are compared to the densities inferred by the DMS in shot 46769. Both discharges are fuelling scans in the Super-X divertor, with the main difference being that the density is increased in 4 steps in 46893, while it is continuously increased in 46769. The 20 lines of sight of the spectrometer are shown in figure \ref{fig:fig18_DMS}, together with the line-averaged densities measured by the DMS and the CIS. The CIS densities have been obtained by modeling the line-integrated spectra over the DMS lines of sight and then fitting the resulting spectra for the most likely electron density and neutral temperature. 
The agreement is reasonable but the line-averaged densities inferred by the CIS tend to under-estimate the DMS values. This could be due to the CIS profile being over-smoothed, thus reducing the peak density along the separatrix, or due to the CIS underestimating the density in regions of low density ($\leq$ 1 $\cdot$ 10$^{19}$ m$^{-3}$), where the CIS measurements are not accurate enough to give precise density measurements.

\section{Discussion}
In the previous sections, the application of a non-linear tomographic inversion on CIS contrast measurements to infer 2D electron and neutral temperature profiles has been described. By measuring the contrast at three different interferometric delays, the contributions of Doppler broadening and Stark broadening can be distinguished. 
\subsection{Analysis improvements}
More complex lineshape models can be included in the analysis, for example allowing for two neutral populations with different temperatures. This could allow to measure the presence of a "hot" and "cold" neutral component, similar to what is observed on TCV \cite{specTCVEPS}. The presence of hydrogen contamination could also be included as a space-dependent parameter, leading to an inferred 2D hydrogen fraction profile. Synthetic testing could be used to test if measurements at three delay values are enough to correctly infer a higher number of parameters than the three used in this work. 

An alternative to increasing the number of delays could be to constrain some of the profiles with data from other diagnostics. For example, including the CIS analysis in an integrated data analysis framework \cite{bowman_development_2020} with the rest of the MWI imaging channels could allow constraining the electron density profile, while adding information on the neutral temperature and hydrogen fraction to the integrated analysis. 

Possible improvements to the optimization algorithm include the addition of physical constraints, such as requiring a profile with a single peak across the separatrix. These kinds of constraints would be important to improve the inference in regions with little or no emission, which are currently unconstrained and just masked off. 

A limitation of the diagnostic is that the density inference becomes more challenging in high-temperature regions, as the broadening becomes completely dominated by the Doppler component. This makes the MAST-U Super-X divertor the perfect environment for this analysis, due to the low electron temperatures involved (0.2 - 4 eV), but also suggests that applying to hotter regions such as close to the X point, might yield worse results. Measuring a higher n Balmer line, such as $D_\delta$ or $D_\epsilon$, would increase the relative magnitude of the Stark component, but it would also lead to a lower signal (factor $\sim 4$ to $\sim 16$ respectively) and thus lower maximum framerate.  For this reason, it may be valuable to complement this kind of analysis with 2D density inferences based on He line ratios, which have been shown to work well at higher temperatures ($\sim$ $\geq$ 10 eV) but have a worse performance at lower temperatures \cite{linehan_validation_2023}. Having shown the feasibility of the technique, further analysis will be performed in future work on the data acquired in the second MAST-U campaign to compare the electron density and neutral temperature profiles in conventional, elongated, and Super-X divertor configurations in different experimental conditions.

\subsection{Physics interpretation of the neutral temperature}
The interpretation of the temperature parameter describing the Doppler broadening as a bulk neutral appears reasonable in regions where ionization is dominant, but it does not obviously apply in regions where molecular interactions or electron-ion recombination are dominant. If the excited neutrals are created via molecular interactions, such as molecular assisted dissociation (MAD), the Doppler broadening could be parameterized by the energy released to the neutrals in the dissociation instead of the bulk neutral temperature. Given the low neutral temperatures inferred it looks unlikely for this to be the case, but further modeling would be needed to confirm this. In regions where the excited neutrals are mainly generated through electron-ion recombination, which can become the dominant emission process near the target in strongly detached cases, and the de-excitation happens before the neutrals have the chance to thermalize with the bulk neutrals, then the temperature describing the Doppler broadening may be the ion temperature instead of the neutral temperature. This could be tested by studying the evolution of the neutral temperature profile before and after the electron-ion recombination becomes dominant, as well as comparing the inferred (ion) temperatures with the ones measured at the target by the Retarding Field Analyzer probe, which can measure the ion temperature at the target. 
\subsection{Phase information}
Only the three fringe contrast measurements are used in this analysis, but the Fourier demodulation also yields three phase measurements. Single-delay CIS phase measurements have traditionally been used to measure ion velocities through the shift in the fringe phase caused by Doppler broadening. A novel simultaneous inversion of CIII ion temperatures and velocities is currently in development at MAST-U \cite{Doyle_inversion}. A similar approach could be used to include the phase data in the inversion and infer neutral velocities as well, albeit with the additional complication of neutral flows being unconstrained to follow the magnetic field direction. This would also allow evaluating possible effects of line-integrated Doppler shifts on the lineshape, which are currently neglected but may appear as an additional effective broadening mechanism. Furthermore, having phase measurements at three different delays could allow distinguishing between the effects of Doppler shifts and hydrogen contamination, which would then feedback into the density inference as a self-consistent constraint on the hydrogen fraction. Similarly to the neutral temperature case, care should then be taken in correctly interpreting the measured velocity, which for example may be more indicative of ion velocities instead of neutral velocities in regions dominated by electron-ion recombination, and will require careful modeling.
\subsection{Outlook}
Having shown the feasibility of the technique, further analysis will be performed in future work on the data acquired in the second MAST-U campaign to compare the electron density and neutral temperature profiles in conventional, elongated, and Super-X divertor configurations. These measurements will allow detailed comparisons with SOLPS-ITER simulations and analytical models, testing our ability to model the divertor behaviour and reducing uncertainties when scaling predictions to reactor-size devices.

\section{Conclusions}
The first measurements of 2D electron density and neutral temperature profiles in the MAST-U divertor with a multi-delay coherence imaging diagnostic have been presented.  
Testing the non-linear tomographic inversion on synthetic measurements has shown that typical errors are distributed in the range [15 \%, 45 \%], with larger errors in regions of low emission or close to the target. Furthermore, reasonable estimations of the errors can be made with a Monte Carlo approach, with the exception of regions at the edge of the view, where even though the uncertainty is correctly estimated to be high ($\geq$ 100 \%), it still underestimates the error. Experimental measurements of electron density show typical densities of $\sim 3 \cdot 10^{19}$ m$^{-3} $ and are in reasonable agreement with the divertor Thomson scattering. Neutral temperatures of $\sim$ 2-3 eV are measured upstream, dropping to low values close to the target, below the sensitive range of the instrument. The typical inferred uncertainties are in the range [20 \%, 50 \%] and account for noise in the contrast images, calibration uncertainties, and uncertainty in the input 2D emissivity profiles, electron temperatures, and polarization angle. 

\section{Acknowledgements}
This work has been carried out within the framework of the EUROfusion Consortium, partially funded by the European Union via the Euratom Research and Training Programme (Grant Agreement No 101052200 — EUROfusion), and from EPSRC Grant EP/S022430/1. The Swiss contribution to this work has been funded by the Swiss State Secretariat for Education, Research and Innovation (SERI). Views and opinions expressed are however those of the author(s) only and do not necessarily reflect those of the European Union, the European Commission or SERI. Neither the European Union nor the European Commission nor SERI can be held responsible for them.

\appendix
\section{Simplified forward models} \label{sec:appendix}
\begin{figure}[h!]\centering
\includegraphics[width= 0.49\textwidth]{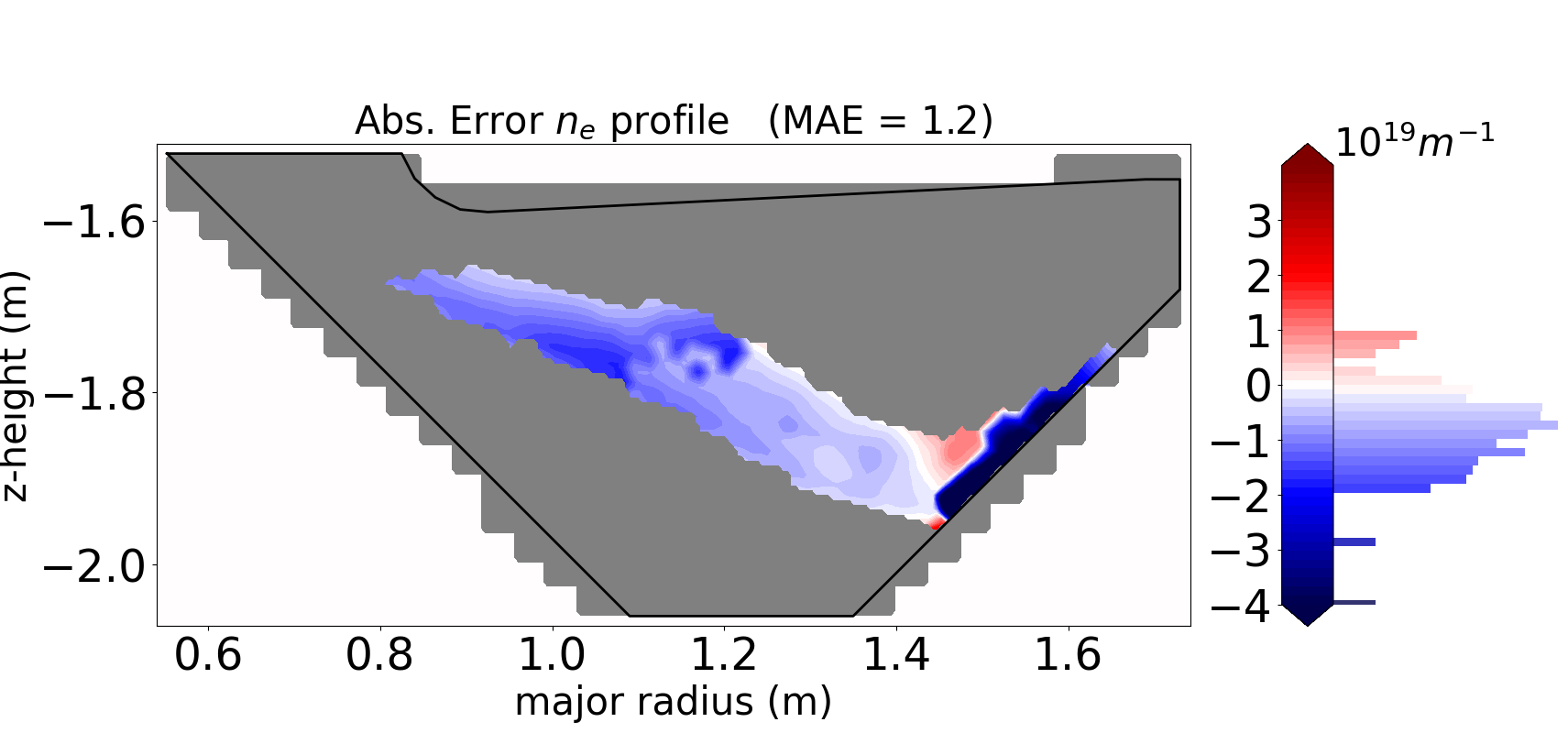}
\includegraphics[width= 0.49\textwidth]{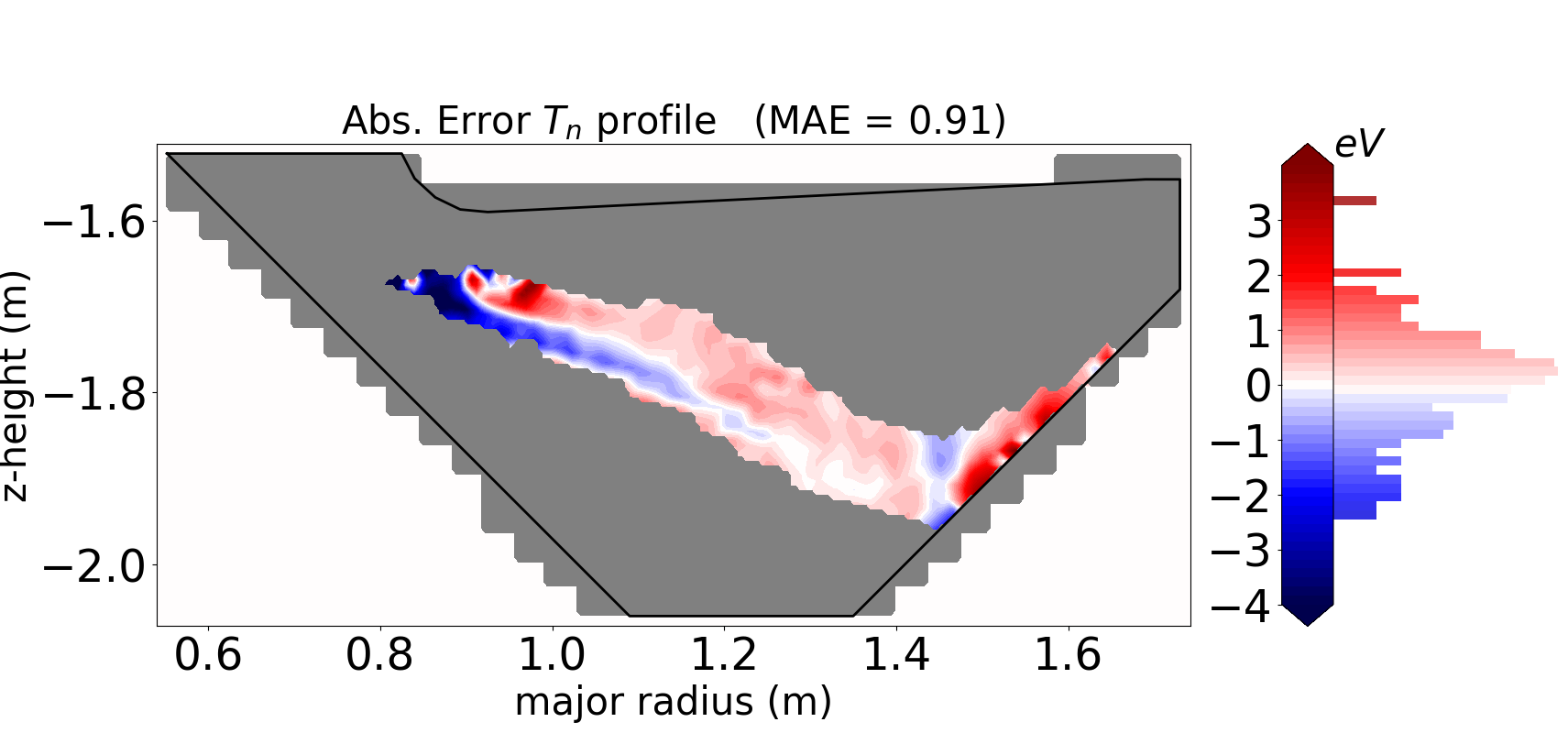}
\vspace{-0.0 cm}\caption{Error in the inferred 2D profiles when the effect of Zeeman splitting is not accounted for in the inversion}
\label{fig18:neglect_Zeeman}
\vspace{-0.3cm} 
\end{figure}

The full lineshape model presented in section \ref{sec:lineshape} can be simplified if faster inversions are required. For example, it could be useful to obtain rough results quickly to inform the choice of parameters for experiments run in succession. A first simplification consists in neglecting the effect of Zeeman splitting that, as shown in figure \ref{fig:fig3_delay_curve_background}, acts as a minor correction in the small magnetic fields of the MAST-U divertor. This allows the inversion to run $\sim$ 3 times faster. The effect of this approximation can be determined by generating synthetic contrast images that include the effect of Zeeman splitting, but neglecting it in the inversion. The resulting error in the density and temperature profiles for the same Super-X SOLPS simulation used in section \ref{sec:Synthetic_test} is shown in figure \ref{fig18:neglect_Zeeman}.  
\begin{figure}[h!]\centering
\includegraphics[width= 0.49\textwidth]{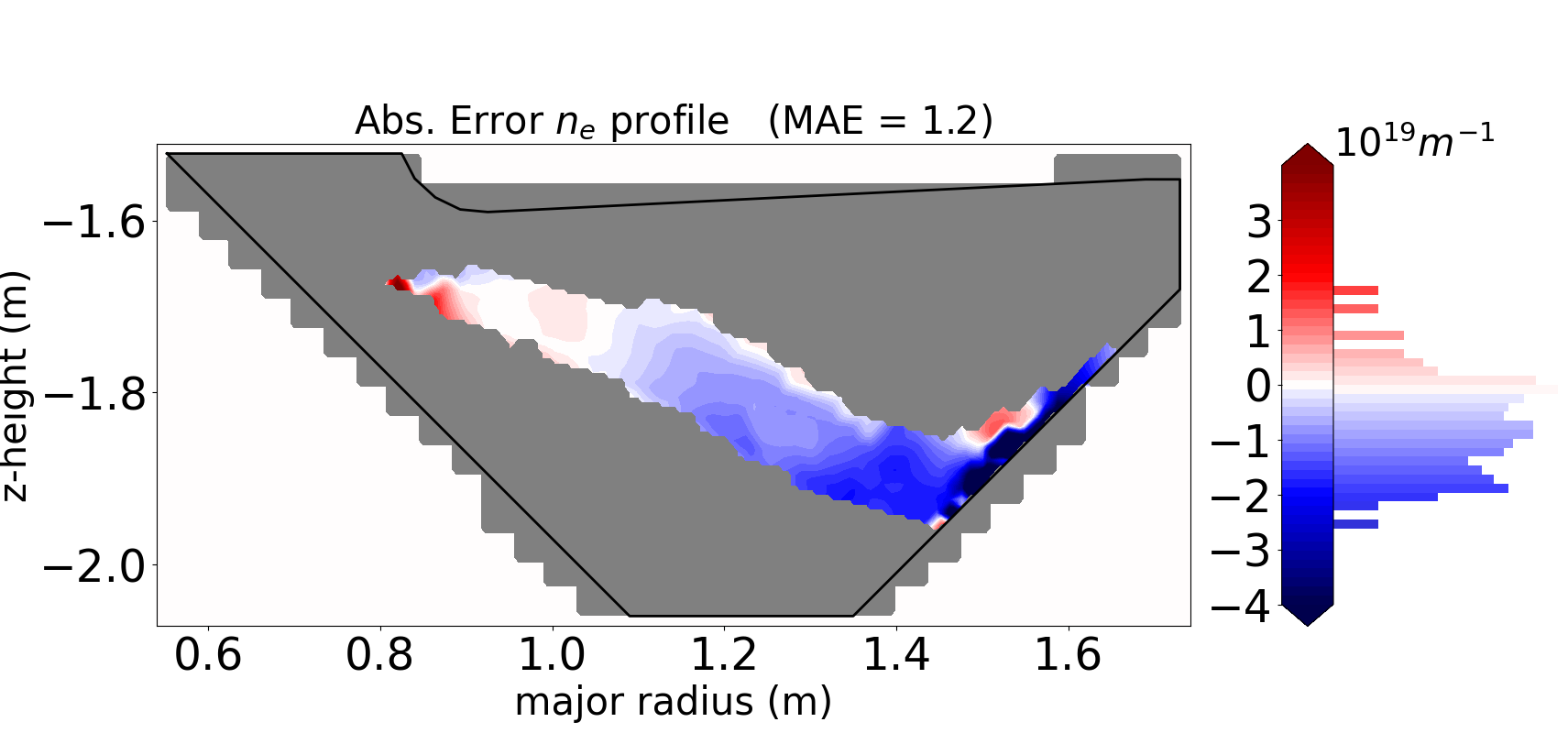}
\includegraphics[width= 0.49\textwidth]{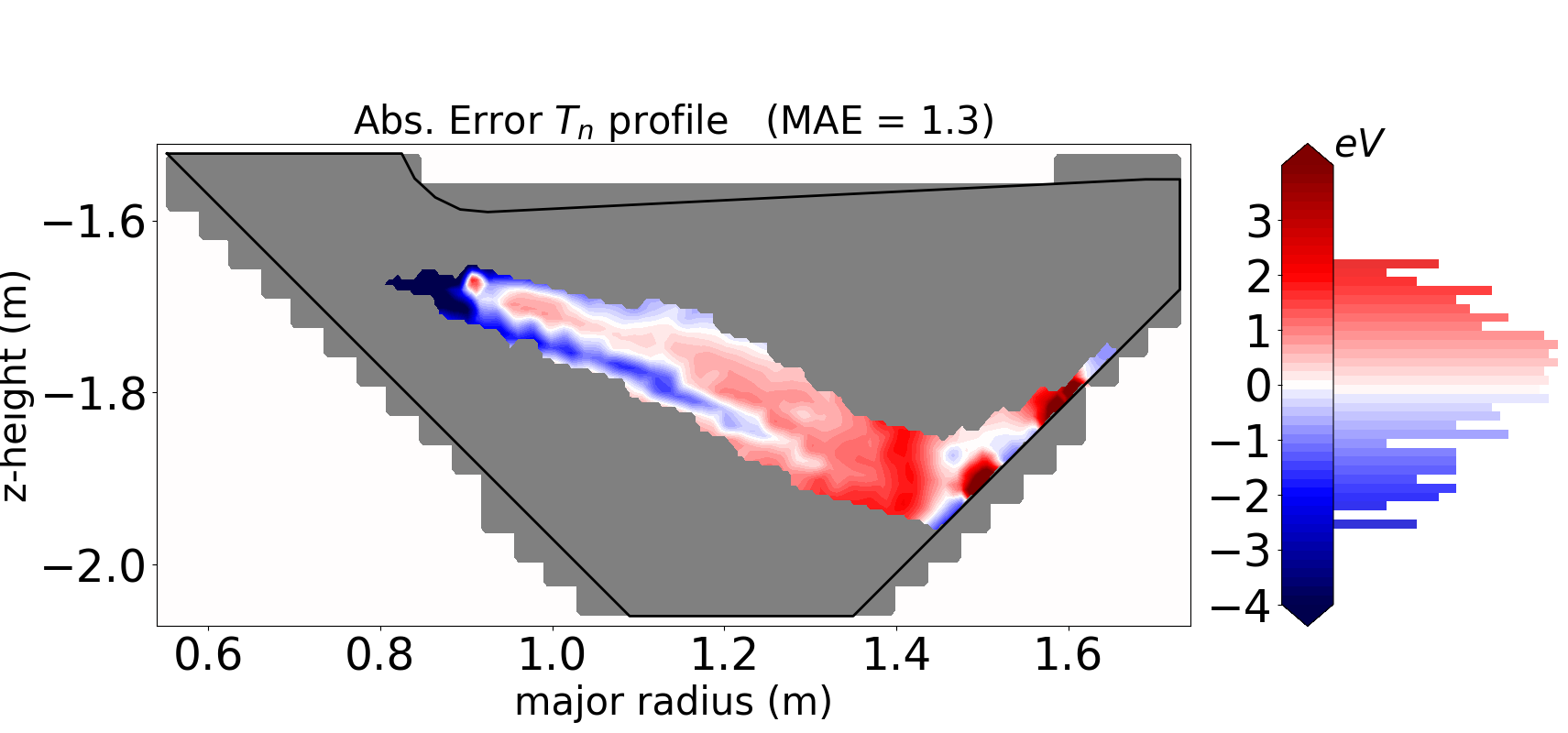}
\vspace{-0.0 cm}\caption{Relative error in the inferred 2D profiles when the delay shear across the sensor is not accounted for in the inversion}
\label{fig19:neglect_delay_shear}
\vspace{-0.3cm} 
\end{figure}
Neglecting Zeeman splitting changes how the broadening is separated between Stark and Doppler broadening. This causes the temperature to become consistently overestimated across the entire profile, while the electron density becomes consistently underestimated. This can be seen as a shift of the bulk of the corresponding error distributions. The mean absolute errors across the masked profiles increase to 1.1 $\cdot 10^{19} m^{-3}$ and 0.93 eV for the electron density and neutral temperature profile respectively. The mean absolute relative errors reach 47 \% and 30 \% respectively, while the median absolute relative errors increase to 44\% and 9.7\%. 
Another possible simplification consists in neglecting the variation of the interferometric delays  across the sensor in the forward model. Along with the assumption that Doppler shifts do not affect the contrast, this allows computing a 2D contrast profile before the line integration, which can then be integrated linearly. Exchanging the matrix product of the geometry matrix and the spectra matrix with the product of the geometry matrix and the contrast vector allows inversions that run $\sim$ 100 times faster. Instead of assuming that all pixels measure the contrast at the same delay as in the center of the sensor, the effect of the assumption can be minimized by taking the average delay in the masked region of the image that is then used in the inversion. The effect of this assumption is shown in figure \ref{fig19:neglect_delay_shear}, where the  errors in the inferred density and temperature profiles are shown. 
In this assumption, the error distribution of both profiles becomes broader. The bulk of the density error distribution shifts to negative values while the temperature error distribution shifts to positive values. The increased error due to this approximation is not distributed uniformly along the spatial profile and it becomes more significant toward the target. This is due to the information regarding the target begin localized mostly in the left edge of the image and the pixels further away from the center of the image being most affected by the approximation. The mean absolute errors increase to 1.2 $\cdot 10^{19} m^{-3}$ and 1.3 eV.  While this approximation does lead to a non-negligible increase in error, this may be offset by trading off the significant decrease in computational time with more complex optimization procedures or by including information from additional diagnostics in the inversion procedure.   

\section{Performance of the uncertainty inference} \label{sec:appendix_error}
\begin{figure}[h!]\centering
\includegraphics[width= 0.49\textwidth]{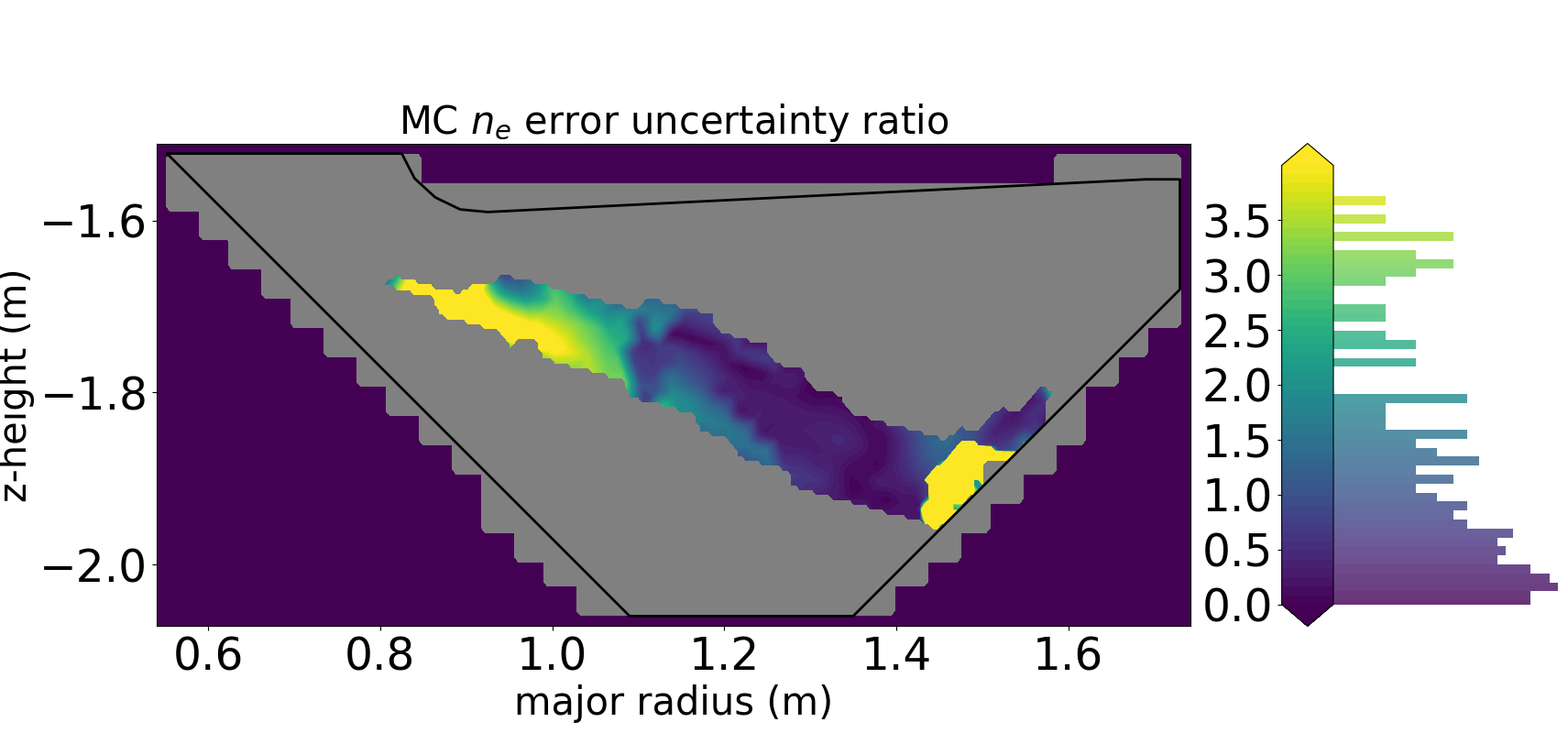}
\vspace{-0.0 cm}\caption{Ratio between error in the inference and estimated uncertainty. Masked-off region with emissivity above 5 \% of the maximum emissivity.}
\label{fig:figB_synth_uncert}
\vspace{-0.3cm} 
\end{figure}

The ratio between the error in the inference and the estimated uncertainty, shown in figure \ref{fig:figB_synth_uncert}, can be used as a metric to check the performance of the uncertainty estimation.

In the assumption of gaussian probability distributions for the inferred density at each grid vertex, the expected distribution in the ratio between the error and the uncertainty would be a gaussian centered in 0 and with a full width half maximum of 1. While the uncertainty is correctly estimated in most of the profile, it can be significantly underestimated at the top left edge of the profile and near the target. The former is attributed to the low electron densities and high neutral temperatures in the region which complicate the density inference and result in an inferred relative uncertainty of 100 \%. The latter is instead due to a very low inferred uncertainty near the target and it is considered a limit of the camera view. 
\section*{References}
\bibliographystyle{iopart-num}
\bibliography{bib}

\end{document}

%% file: packages_commands.tex
 \usepackage{hyperref}
 \usepackage{color}
 \usepackage[utf8]{inputenc}
 \usepackage{bm}
 \usepackage{graphicx, subcaption}
 \usepackage{latexsym}
 \usepackage{pifont}
 \usepackage{amssymb}
 \usepackage{hyperref}
 \usepackage{titlesec}
 \expandafter\let\csname equation*\endcsname\relax
 \expandafter\let\csname endequation*\endcsname\relax
 \usepackage{amsmath}
 \usepackage{physics}
 \usepackage{listings}
 \usepackage{verbatim}
 \usepackage{geometry}
 \usepackage{booktabs}
 \usepackage{enumitem}
 \usepackage[normalem]{ulem} 
 \usepackage{breqn}
 \usepackage[toc,page]{appendix}
 
 \definecolor{red}{rgb}{1.0,0.0,0.0}
 \definecolor{gre}{rgb}{0.0,1.0,0.0}
 \definecolor{blu}{rgb}{0.0,0.0,1.0}

 \definecolor{ora}{rgb}{1.0,0.5,0.0}
 \definecolor{gra}{rgb}{0.0,0.5,1.0}

 \newcommand{\ba}{\begin{abstract}}
 \newcommand{\ea}{\end{abstract}}
 \newcommand{\be}{\begin{equation}}
 \newcommand{\ee}{\end{equation}}

 \newcommand{\rom}[1]{\uppercase\expandafter{\romannumeral #1 \relax}}